\documentclass{jfm}
\usepackage[table,xcdraw]{xcolor}
\usepackage{algorithm}
\usepackage[noend]{algpseudocode}
\usepackage{graphicx}
\usepackage{amsmath}
\usepackage{amssymb}
\usepackage[utf8]{inputenc}
\usepackage{multirow}
\usepackage{booktabs}
\usepackage{array}
\usepackage{amsfonts}
\usepackage{newtxtext}
\usepackage{newtxmath}
\usepackage{natbib}
\usepackage{hyperref}

\renewcommand\Rey{\mbox{\textit{Re}}}  
\renewcommand\Pen{\mbox{\textit{Pe}}}  
\algrenewcommand\algorithmiccomment[1]{\hfill$\triangleright$ #1}
\usepackage{hyperref}
\title{
Effects of Wall Roughness on Coupled Flow and Heat Transport in Fractured Media}

\author{Alessandro Lenci\aff{1,2,3}
  \corresp{\email{alessandro.lenci@unibo.it}},
    Yves Méheust\aff{3,4},
  Maria Klepikova\aff{3},
    Vittorio Di Federico\aff{1},
 \and Daniel M. Tartakovsky\aff{2}}

\affiliation{
\aff{1}Department of Civil, Chemical, Environmental, and Materials Engineering, Università di Bologna, Viale del Risorgimento 2, Bologna 40126, Italy
\aff{2}Department of Energy Science and Engineering, Stanford University, Stanford, CA 94305, USA
\aff{3}University of Rennes, CNRS, Géosciences Rennes -- UMR 6118, F-350042 Rennes, France
\aff{4}Institut Universitaire de France (IUF)}

\begin{document}
\maketitle

\begin{abstract}
Heat transfer in fractured media results from the interplay between advective transport within the fracture and conductive heat exchange with the surrounding rock matrix. Aperture heterogeneity structures this interplay by generating preferential flow channels and quasi-stagnant zones, leading to early-time anomalous transport dominated by advective channeling and to late-time non-Fickian dynamics controlled by matrix conduction. This study develops a physics-based stochastic framework that couples a time-domain random walk (TDRW) representation of in-fracture advection and conduction with a semi-analytical description of matrix--fracture heat exchange, enabling a unified characterization of both short- and long-time anomalous heat-transport regimes. Matrix trapping times follow a L\'evy--Smirnov distribution derived from first-passage theory, and the interfacial heat flux is evaluated through a nonlocal Duhamel kernel that rigorously captures the temporal nonlocality imposed by heat-conduction theory. Monte Carlo simulations over stochastic aperture fields elucidate the roles of fracture closure, correlation length, and P\'eclet number in shaping heat transport. Increasing fracture closure enhances channelization and accelerates early-time heat transport, whereas larger correlation lengths amplify anomalous spreading. Higher P\'eclet numbers strengthen advective dominance but do not suppress the long-time subdiffusive tail induced by matrix conduction. Breakthrough curves exhibit heavy-tailed decay consistent with Lévy–Smirnov trapping induced by semi-infinite matrix diffusion. Results reveal a transition from superdiffusive to subdiffusive transport governed by advective channeling, aperture-induced dispersion, and matrix conduction. The framework provides a predictive and computationally efficient route for modelling heat transport in heterogeneous fractures, with relevance to geothermal energy extraction, subsurface thermal storage, and engineered thermal systems.
\end{abstract}


\section{Introduction}\label{sec:intro}
Quantitative understanding of heat and mass transport in heterogeneous porous media remains a fundamental challenge across numerous scientific and engineering disciplines. Applications as diverse as the use of heat as tracer to characterize heterogeneous (including fractured) subsurface media \citep{sillimanGroundwater1989,klepikovaJHydrol2011}, 
thermal remediation of polluted soils \citep{hinchee1992BOOK}, 
geothermal energy \citep{shaikApplThermalEng2011} 
and blood perfusion \citep{pennes1948} all 
involve intricate interactions between advective transport through high-permeability pathways and diffusive exchange with surrounding low-permeability matrices. These systems are typically characterized by structural heterogeneity spanning many orders of magnitude in scale, resulting in strongly non-Fickian behavior, e.g., long-tailed residence time distributions \citep{roubinet-2013-particle}. 

The exchange between mobile and immobile zones in porous media, e.g., fractures and matrix and/or preferential flow paths and stagnant domains, controls large-scale transport efficiency, long-term retention, and the evolution of reactive processes \citep[and references therein]{yang-2025-dual}. Development of robust modelling frameworks that accurately capture these multiscale, flow-driven exchange mechanisms is essential for predictive analysis, risk assessment, and optimal design in a plethora of applications. In the case of heat transport, this exchange corresponds to conductive heat transfer from fractures into the surrounding rock matrix and heat transport within that matrix.

The present work addresses coupled flow and heat transport in fractured geological formations, which are a good framework to illustrate the modelling challenges. Subsurface environments are often composed of a low-permeability porous matrix (granite, shale, or metamorphic rock, etc.), with intrinsic permeability typically in the $10^{-20}$–$10^{-16}~\mathrm{m}^2$ range.
The matrix contains a network of interconnected fractures with local permeability as high as $10^{-10}~\mathrm{m}^2$, depending on fracture aperture, spacing, orientation distribution, and degree of connectivity \citep{bonnetRevGeophys2001,viswanathan-2022-from}. This contrast in hydraulic properties restricts fluid flow to the fractures, while heat transport in the surrounding matrix is dominated by heat conduction. 

Fracture--matrix interactions control the partitioning of heat between mobile and immobile domains and therefore the degree to which the surrounding rock participates in thermal storage \citep[and references therein]{ruiz-2014-analytical}. Even modest geometric variability within a single fracture can reorganise the flow into preferential channels and low-velocity pockets, producing sharp thermal fronts and spatially uneven heat uptake \citep{Gisladottir2016}. 
Numerical analyses of heat transfer in discrete–fracture networks \citep[e.g.][]{Geiger2010a} and in explicitly resolved rough-walled single fractures \citep[e.g.][]{Klepikova2021} show that structural heterogeneity, arising respectively from network connectivity or from aperture variability, organises the flow into preferential channels and regions of quasi-immobile flow.  These heterogeneous flow patterns enhance advective intermittency and promote repeated equilibration with the surrounding rock matrix, thereby exerting a strong control on the timing of thermal breakthrough, thermal recovery, and overall heat extraction efficiency.

Field-scale thermal tracer tests \citep{Klepikova2016, LaBernardie2018} exhibit analogous behaviour: early breakthrough, a gradual intermediate decline, and tails, consistent with the combined effects of channelised flow, matrix conduction, and intermittent trapping. Together, these observations demonstrate that realistic fracture geometries govern the large-scale thermal response of fractured media and motivate modelling strategies that explicitly resolve aperture variability rather than relying on homogenised representations.

Furthermore, at the fracture scale, geological fracture walls exhibit self-affine roughness \citep{Schmittbuhl1995a} and are unmatched below a characteristic correlation length \citep{Brown1995}. This generates heterogeneous aperture fields containing nearly closed zones as well as spatially correlated large-aperture channels extending up to the scale of the correlation length. The resulting flow patterns are strongly channelised, with channeling intensity all the larger as the fracture is more  closed (i.e., as aperture heterogeneity is larger \citep{Brown1987,Meheust2001}), and with preferential flow-channel sizes (as well as low permeability region sizes) limited by the underlying correlation scale \citep{Meheust2003,Lenci2022a}. Consequently, fracture flow can deviate markedly from parallel-plate predictions based solely on the mean aperture.

This geometric variability dictates thermal breakthrough behaviour through its control on the competition between advective transport along preferential channels and diffusive exchange in low-velocity pockets \citep{Neuville2010,Klepikova2021}. Across both network and single-fracture scales, the coexistence of preferential conduits and quasi-immobile zones causes transport to alternate between advection- and diffusion-dominated regimes, producing strong spatial intermittency. Efficiently capturing this behaviour requires mesh-free particle-based methods capable of resolving the broad spectrum of residence times induced by roughness-controlled channeling.

We present here a hybrid stochastic–physical model for heat transfer in a fracture-matrix system, which accounts both for both heat advection within fractures, heat conduction in the fluid, and exchange of heat with a heat-conducting matrix. Heat advection and conduction in the flowing fluid are modeled with a time-domain random walk (TDRW) scheme \citep{Delay2001,Russian2016}, i.e., heat particle trajectories in the fracture evolve through random time increments while spatial displacements are dictated by the local depth-averaged velocity. This framework efficiently captures the strong velocity contrasts of rough fractures: particles propagate rapidly along high-velocity channels and remain effectively immobilised in nearly closed regions, without requiring prohibitively small time steps or finely resolved grids \citep{Noetinger2016}. The fracture–matrix exchange 
is modeled by an analytically derived kernel: the matrix residence-time distribution is obtained directly from the Green’s function of transient one-dimensional conduction, yielding a calibration-free and thermodynamically consistent description of conductive heat loss from the fracture. Stochastic propagation is performed over explicitly resolved rough-walled geometries so that aperture-controlled heterogeneity directly determines both advective residence times and local exchange histories. We thus obtain a predictive and computationally efficient representation of non-Fickian thermal transport that retains the full imprint of the heat transport processes at play.

The model is validated against analytical solutions and high-fidelity finite-element simulations. To systematically assess the influence of heterogeneity and flow conditions, we conduct a Monte Carlo analysis over multiple combinations of relative fracture closure, spatial correlation length, and Péclet number, thereby encompassing a broad range of geologically plausible scenarios. Results are presented in terms of four key observables: mean longitudinal displacement, displacement variance, breakthrough curves, and heat–exchange efficiency at the fracture–matrix interface. These quantities allow the distinction among transport regimes (ballistic, diffusive, subdiffusive), characterise the spreading of the thermal front, and quantify the efficiency of heat extraction over time. Their temporal evolution provides a concise macroscopic signature of how advective and conductive mechanisms jointly shape heat transport in rough–walled fractures, complementing the process-level understanding derived from the underlying stochastic model.

The manuscript is organised as follows. 
Section~\ref{sec:medium} introduces the geometrical characterisation of the fracture aperture field and the generation of synthetic rough-walled fractures. 
Section~\ref{sec:flow} presents the depth-averaged flow model based on the lubrication approximation, and Section~\ref{sec:transport} details the time-domain random walk (TDRW) formulation coupled with the semi-analytical fracture--matrix heat-exchange model. 
Validation against analytical solutions and finite-element simulations is provided in Section~\ref{sec:validation}. 
Section~\ref{sec:analysis} examines the statistical behaviour of heat transport across varying heterogeneity levels and Péclet numbers, including displacement moments, breakthrough curves and heat exchange efficiency. Conclusions and perspectives are summarised in Section~\ref{sec:conclusion}. Appendix~\ref{appA} derives the matrix trapping-time distribution, Appendix~\ref{appB} presents the associated memory kernel, and Appendix~\ref{appC} outlines numerical implementation aspects.

\section{Synthetic Geological Fractures}\label{sec:medium}

Geological fractures are discontinuities in the rock matrix, which arise from fracturing in the material due to tectonic or thermal constraints, and which are quasi-planar at large scales. They are inherently heterogeneous. Laboratory and field measurements show that the roughness of both synthetic \citep{SchmittbuhlJGR95} and natural \citep{Renard2013} fracture surfaces displays self-affine scale-invariance over several orders of magnitude. When data obtained from different measurement techniques and surfaces of different sizes are combined, the scaling extends over 7--12 decades, from the smallest resolved laboratory scales up to the dimensions of large faults \citep{Candela2009}. This self-affinity means that the surfaces are statistically invariant by any rescaling of the in-plane coordinates by any factor $\lambda$, provided that the out-of-plane coordinate is rescaled by $\lambda^H$, the so-called {\em Hurst} exponent being primarily characteristic of the fracturing process. The value of this exponent is independent of rock type (and even, of brittle material type) or geologic and tectonic environments \citep{Bouchaud1990,Renard2013,Milanese2019}. An exception arises in sandstones, where intergranular fracturing sets the lower scale for self-affinity to the grain scale, and induces a lower Hurst exponent, with $H \approx 0.45$ \citep{Boffa1999}. Measurements on exposed fault surfaces indicate the combined action of brittle failure, plastic deformation, and three-body wear leads to a universal smoothing process \citep{Sagy2007}, which may alter the roughness amplitude in time (in particular as a consequence of fault slip), but conserves the surfaces' self-affinity. However, recent numerical results show that self-affine walls may also develop from initially flat surfaces due to three-body wear over large times \citep{Milanese2019}.
In Fourier space, the self-affine scale-invariance of the fracture walls is reflected in a power-law decay of the power spectral density (PSD) of their topographies,
\begin{equation}\label{eq:PSD_kappa}
    \mathcal{F}(\kappa) \propto \kappa^{-(2+2H)},
\end{equation}
which holds for wavenumbers $\kappa$ between cutoffs associated to the
aforementioned characteristic scales. The wavenumber provides a natural
indexing of the spatial scales of the roughness: modes with small
$\kappa$ correspond to large-scale aperture variations, whereas modes
with large $\kappa$ represent fine-scale variations.

\begin{figure}
\centerline{\includegraphics[width=1\textwidth]{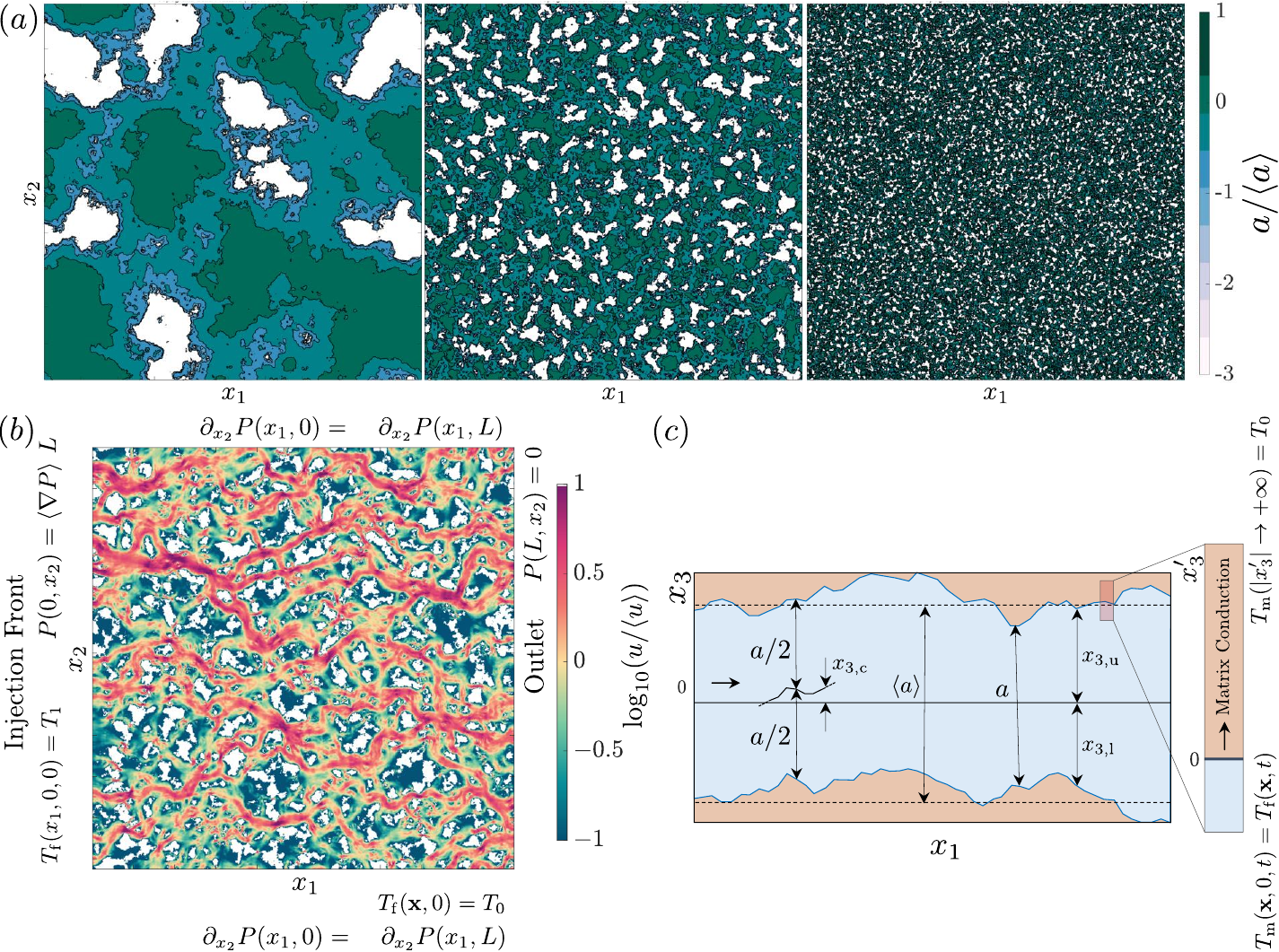}}
\caption{(a) Synthetic fracture aperture fields for increasing correlation ratios $L / L_\mathrm{c} = \{2,\,16,\,64\}$ (left to right), all with the same fracture closure $\sigma_a / \langle a\rangle = 0.8$. (b) Dimensionless velocity magnitude field, $\log_{10}(u/\langle u\rangle)$, for the case $L / L_\mathrm{c} = 16$, showing the flow structure under the applied boundary conditions. (c) Transverse fracture profile illustrating wall roughness, aperture geometry, matrix domain, and the initial and boundary conditions for flow and transport. Panels (b) and (c) also indicate the transport mechanisms considered in both fracture and matrix. All synthetic aperture fields were generated using $L_\mathrm{c} = 0.1~\textrm{m}$, $H = 0.8$, $a_\mathrm{m} = 1~\textrm{mm}$, and $\sigma_a / \langle a \rangle = 0.8$.}
\label{Fig1}
\end{figure}

The separation between the two rough walls of a fracture defines the fracture's aperture field (Fig.~\ref{Fig1}c). For a horizontal fracture with upper and lower walls of respective topographies $x_{3,\mathrm{u}}$ and $x_{3,\mathrm{l}}$:
\begin{equation}\label{eq:aperture_definition}
    a(\mathbf{x}) = x_{3,\mathrm{u}}(\mathbf{x}) - x_{3,\mathrm{l}}(\mathbf{x}) + a_\mathrm{m},
\end{equation}
where $\mathbf{x}=(x_1,x_2)^\top$ is the in-plane position vector, and $a_\mathrm{m}$ is the mechanical aperture, defined as the distance between the (parallel) mean planes of the opposing walls. Surface features of the two walls tend to match above a characteristic in-plane scale, the correlation length $L_\mathrm{c}$  \citep{Brown1995,Meheust2003}. Hence, for wavelengths smaller than $L_\mathrm{c}$ (i.e., $\kappa > 2\pi/L_\mathrm{c}$), the self-affine wall topographies, which are uncorrelated with each other at these scales, provide the aperture field with the same self-affinity, so that its PSD exhibits the same characteristic power-law behavior \eqref{eq:PSD_kappa} \citep{Brown1995}. In contrast, at length scales larger than $L_\mathrm{c}$ (i.e., $\kappa < 2\pi/L_\mathrm{c}$), the PSD flattens. The fracture aperture is thus a spatially heterogeneous random field characterized by its mean aperture $\langle a \rangle$, standard deviation $\sigma_a$, characteristic length $L_\mathrm{c}$ controlling the scale range for self-affinity, fracture size $L$ setting the domain extent, and Hurst exponent $H$. 

Synthetic aperture fields with prescribed statistical properties are generated using fast Fourier transform (FFT)–based spectral synthesis \citep{Meheust2003,Lenci2022a}. For a given roughness amplitude $\sigma_a$, the degree of fracture closure is controlled by adjusting the mean aperture $\langle a\rangle$. Where the resulting aperture becomes negative, corresponding to wall–wall contact, the aperture is set to zero, yielding a connected network of flow paths interspersed with impermeable contact zones \citep{Brown1995,meheustJGR2001}. If no contact occurs, the mean aperture coincides with the mechanical aperture, $\langle a\rangle = a_\mathrm{m}$; otherwise it exceeds it. Representative realizations are shown in Fig.~\ref{Fig1}a for correlation ratios
$L/L_\mathrm{c}=\{2,16,64\}$ and fracture closure $\sigma_a/\langle a\rangle=0.8$.

The aperture fields are generated directly on the computational grid, with the self-affine power spectrum sampled over the range of wavenumbers resolved by the mesh, thereby avoiding interpolation or mesh refinement. Although self-affinity formally implies increasing aperture gradients as the grid spacing decreases, the amplitude of the added high-wavenumber modes simultaneously vanishes, so that mesh refinement does not significantly affect the validity of the lubrication approximation in practice. The grid resolution is therefore chosen to resolve flow structures well below the correlation length while keeping aperture gradients within acceptable lubrication limits over most of the domain. Under this criterion, both the statistical properties of the aperture field and the resulting hydraulic and thermal responses are insensitive to moderate changes in grid resolution.

Note that the heterogeneity of the aperture field is jointly controlled by the fracture closure ratio, $\sigma_a / \langle a\rangle$,  the correlation ratio, $L / L_\mathrm{c}$, and the Hurst exponent. The fracture closure prescribes the relative amplitude of aperture fluctuations with respect to a given mean aperture, thereby setting the overall contrast of the local transmissivity 
field and controlling in particular the probability density function (PDF) of local apertures.
 The correlation ratio $L/L_{\mathrm{c}}$ 
determines how 
many statistically distinct roughness domains are represented within a single 
realisation. Hence, large $L/L_{\mathrm{c}}$ values correspond to aperture fields composed of many quasi-independent geometric segments, whereas small $L/L_{\mathrm{c}}$ indicates 
that the geometry is dominated by a few long-wavelength features. 
Finally, the Hurst exponent $H$ specifies the degree of self-affine smoothness of the walls, controlling the relative weight of long- versus short-wavelength asperities in the aperture spectrum. 

The thermal properties of water and granite (Table~\ref{Tab1}) are taken from standard datasets \citep{Incropera2007} and from measurements in crystalline formations \citep{Klepikova2016, Kant2017}, which ensures consistency with typical hydrothermal conditions. The geometric parameters of the medium (Hurst exponent, correlation ratio, and closure ratio) follow the choices justified in Section~\ref{sec:medium} based on documented statistical features of natural self-affine fractures. The numerical aspects related to aperture-field generation, mesh discretisation, and solution of the lubrication flow follow the implementation detailed
in \citet{Lenci2022a}.

\begin{table} 
\centering
\caption{Physical and numerical parameters defining the fracture geometry, 
fluid properties, and matrix characteristics. 
All values are fixed across all simulations.}
\label{Tab1}
\begin{tabular}{@{}lll@{}}
\toprule
\textbf{Category} & \textbf{Parameter} & \textbf{Value} \\
\midrule
Fracture 
    & $\langle a\rangle$     & $10^{-3}~\mathrm{m}$ \\
    & $L_\mathrm{c}$         & $0.1~\mathrm{m}$ \\
    & $H$                    & $0.8$ \\
\midrule
Fluid 
    & $\mu_\mathrm{w}$       & $10^{-3}~\textrm{Pa}\cdot\textrm{s}$ \\
    & $\rho_\textrm{w}$      & $1000~\textrm{kg}/\mathrm{m}^3$ \\
    & $k_\mathrm{w}$         & $0.59~\textrm{W}/(\mathrm{m}\cdot\textrm{K})$ \\
    & $c_{p,\mathrm{w}}$     & $4189~\textrm{J}/(\textrm{kg}\cdot\textrm{K})$ \\
    & $D_\mathrm{f}$         & $1.41\times 10^{-7}~\mathrm{m}^2/\textrm{s}$ \\
\midrule
Matrix
    & $\rho_\textrm{r}$      & $2500~\textrm{kg}/\mathrm{m}^3$ \\
    & $k_\textrm{r}$         & $3.5~\textrm{W}/(\mathrm{m}\cdot\textrm{K})$ \\
    & $c_{p,\textrm{r}}$     & $750~\textrm{J}/(\textrm{kg}\cdot\textrm{K})$ \\
    & $\alpha_\mathrm{m}$    & $1.87\times 10^{-6}~\mathrm{m}^2/\textrm{s}$ \\
    & $\phi_\mathrm{r}$      & $0.1$ \\
    & $\phi_\mathrm{m}$      & $0.50$ \\
\midrule
Numerical 
    & $N_\textrm{MC}$        & $10^2$ realizations per combination \\
    & $N_\textrm{p}$         & $10^{6}$ particles per realization \\
    & $N_\textrm{mesh}$      & $2^{10}\times 2^{10}$ finite volumes \\
\bottomrule
\end{tabular}
\end{table}
\section{Flow in Fractures with Self-affine Walls} \label{sec:flow}
%
Consider the flow of an incompressible fluid, with density $\rho_\mathrm{w}$ 
and dynamic viscosity $\mu_\mathrm{w}$, in a horizontal fracture with mechanical aperture $a_\mathrm{m}$. The flow is driven by an imposed pressure gradient $\langle\nabla P\rangle$ and is characterized by the Reynolds number 
\begin{equation} \label{eq:Reynolds_Number}
\Rey = \frac{\rho_\mathrm{w} U_\mathrm{c} \ell_{x_3}^2}{\mu_\mathrm{w}\, \ell_{\mathbf{x}}},
\end{equation}
where $U_\mathrm{c}$ is a characteristic fluid velocity assumed equal to the maximum velocity within a parallel plate fracture of aperture $a_\mathrm{m}$, that is, $u_\textrm{max}=a_\mathrm{m}^2\langle \nabla P\rangle /8\mu_\mathrm{w}$; and $\ell_{x_3} = a_\mathrm{m}$ and $\ell_{\mathbf{x}} = L_\mathrm{c}$ denote characteristic length scales associated with the vertical and in-plane variations of the velocity field, respectively \citep{Meheust2001,Neuville2011}. Under these conditions, the steady-state fluid velocity $\mathbf{u}^\prime(\mathbf{x}^\prime) = (u_1, u_2, u_3)^\top$ and pressure $p(\mathbf{x}^\prime)$ within a fracture $\Omega_\mathrm{f}^\prime = \left\{ \mathbf{x}' = (x_1, x_2, x_3)^\top \in \mathbb{R}^3 : (x_1, x_2) \in (0, L) \times (0, L),\; x_{3,\mathrm{l}} \le x_3 \le x_{3,\mathrm{u}} \right\}
$ satisfy 
the incompressible Stokes equations
\begin{align}\label{eq:3DStokes}
\rho_\mathrm{w} \mathbf{g} - \boldsymbol{\nabla}^\prime p + \mu_\mathrm{w}  \boldsymbol \nabla^{\prime 2} \mathbf{u}^\prime = \mathbf{0}, 
\qquad
\boldsymbol{\nabla}^\prime \cdot \mathbf{u}^\prime = 0;
\qquad \mathbf{x}^\prime \in \Omega_\mathrm{f}^\prime;
\end{align}
where $\mathbf{g} = (0, 0, -g)^\top$ is the gravitational acceleration constant, and the operators $\boldsymbol{\nabla}^\prime$ and $\boldsymbol \nabla^{\prime 2}$ are the gradient and Laplacian in three dimensions.

Randomness of the fracture wall topographies $x_{3,\mathrm{u}}(\mathbf{x})$ and $x_{3,\mathrm{l}}(\mathbf{x})$ translates into randomness of the three-dimensional flow domain $\Omega^\prime_\mathrm{f}$. Hence,~\eqref{eq:3DStokes} is an example of partial differential equations on random domains \citep{Xiu2006}, a class of problems that is notoriously challenging to solve, and computationally intensive.
Instead, we take advantage of the geometry of geological fractures and of typical flow conditions within them. 
Since the fracture walls are impermeable and the aperture field $a(\mathbf{x})$ varies smoothly in
$\Omega_\mathrm{f}$, with $\|\nabla a\|\ll 1$ and $a\ll L_\mathrm{c}$, the flow satisfies the classical lubrication
limit. In this regime, the aperture is sufficiently small compared with the in-plane length scales for
vertical velocity components to be much smaller than the in-plane components, so that $u_3 \ll u_1, u_2$. Velocity variations across the gap dominate those within the fracture plane, implying
$\partial_{x_3}^2\mathbf{u} \gg \partial_{x_1}^2\mathbf{u},\,\partial_{x_2}^2\mathbf{u}$. 
The pressure is uniform across the aperture to leading order and depends only on the in-plane coordinates. Under these assumptions,~\eqref{eq:3DStokes} simplifies to
\begin{equation}\label{eq:Stokes}
\mu_\mathrm{w}\,\frac{\partial^2 \mathbf{u}}{\partial x_3^2} = \boldsymbol{\nabla} P, \qquad \mathbf{x} \in \Omega_\mathrm{f};
\end{equation}
where $\Omega_\mathrm{f} = \{ \mathbf{x}=(x_1,x_2)^\top \in \mathbb{R}^2 : (x_1,x_2)\in(0,L)\times(0,L) \}$
denote the in-plane fracture domain, $\mathbf{u}=(u_1, u_2)^\top$ is the in-plane velocity and $P = p + \rho_\mathrm{w} g x_3$ is the reduced pressure that absorbs the gravitational contribution. 
Integrating Eq.~\eqref{eq:Stokes} twice with respect to $x_3$, and imposing no-slip boundary conditions $\mathbf{u}(x_1,x_2,x_{3,\textrm{c}}\pm a/2) = 0$, where $x_{3,\textrm{c}}=(x_{3,\mathrm{u}}+x_{3,\mathrm{l}})/2$ tracks the mean topography, leads to the local Poiseuille law relating the velocity to the reduced pressure gradient:
\begin{equation}\label{eq:velocity_profile}
\mathbf{u}= -\frac{1}{2\mu_\mathrm{w}} 
\left[ \frac{a^2}{4} - 
\left( x_3 - x_{3,\textrm{c}} \right)^2 \right]\boldsymbol{\nabla} P.
\end{equation}
We define the local flux field $\mathbf{j} = (j_1, j_2)^\top$ as the integral of the velocity profile across the fracture aperture,
\begin{equation}\label{eq:cubic}
\mathbf{j} = \int_{x_{3,\textrm{c}}-a/2}^{x_{3,\textrm{c}}+a/2} \mathbf{u}\, \mathrm{d}x_3 = -\frac{a^3}{12\mu_\mathrm{w}}\, \boldsymbol{\nabla} P.
\end{equation}
Due to the no-slip velocity boundary condition at the walls, the three-dimensional incompressibility condition $\boldsymbol{\nabla}^\prime \cdot \mathbf{u}^\prime = 0$ translates into the two-dimensional incompressibility condition for the volumetric flux, $\boldsymbol{\nabla} \cdot \mathbf{j} = 0$.
Combined with Eq.~\eqref{eq:cubic}, this incompressibility condition yields the Reynolds equation for pressure in variable-aperture fractures,
\begin{equation}\label{eq:Reynolds}
\boldsymbol{\nabla} \cdot ( a^3 \boldsymbol{\nabla} P ) = 0, \quad \mathbf{x} \in \Omega_\mathrm{f}.
\end{equation}
This equation contains a random coefficient, $a(\mathbf{x})$, but is defined on the deterministic domain $\Omega_\text{f}$.  
It is subject to a constant macroscopic pressure gradient of magnitude $\langle \nabla P \rangle = \|\partial_{x_1} P\|$ along the $x_1$-axis. Dirichlet boundary conditions are prescribed such that the pressure decreases linearly along $x_1$, with $P(0,x_2) = \langle \nabla P\rangle L$ and $P(L,x_2) = 0$. Lateral periodic boundary conditions along $x_2$ enforce both pressure and flux continuity, 
i.e. $P(x_1,0)=P(x_1,L)$ and $\partial_{x_2}P(x_1,0)=\partial_{x_2}P(x_1,L)$.

The lubrication formulation adopted here has long been established as a reliable representation of flow in rough-walled fractures, providing a physically consistent depth-averaged equivalent for the Stokes equations at small aspect ratios and under creeping flow conditions (i.e., for Reynolds numbers of order smaller than 1).  
Extensive numerical and experimental benchmarks (\citealt{Mourzenko1995,Zimmerman1996,Neuville2010,Neuville2013}) have shown that the resulting Reynolds equation reproduces transmissivity and pressure distributions with excellent accuracy across a wide range of natural roughness levels, while deviations are confined to near-closure zones that contribute negligibly to the total flux owing to the cubic dependence $q\!\propto\!a^3$.  
Recent analyses (\citealt{Krishna2025}) further confirmed that this depth-integrated framework remains robust even beyond the classical slope limits, making it well suited for the low-Reynolds-number, rough-aperture regime considered here.

In the depth-averaged formulation, the local transmissivity entering the Reynolds equation scales as $a^3$, so geometric variability is mapped nonlinearly into the hydraulic response. The geometric parameters that control the local aperture field therefore play a central role in governing the degree and spatial structure of flow localisation. The fracture closure $\sigma_a/\langle a\rangle$ sets the width of the aperture distribution at fixed mean aperture, and thus directly modulates the overall contrasts in the transmissivity field: small changes in the fluctuation amplitude of $a$ translate into much larger variations in $a^3$. Thus, a higher fracture closure enhances the formation of narrow, high transmissivity pathways. On the other hand, the value of $L/L_{\mathrm{c}}$ controls 
the spatial extent of connected high aperture preferential flow channels and connected low aperture barrier regions (see Section~\ref{sec:medium}), thus controlling the spatial organisation of the depth-averaged velocity field. This appears clearly in Fig.~\ref{Fig1}b for $L/L_\mathrm{c}=16$, where preferential flow channels appear in red and low velocity regions in teal color.



Together, fracture closure and the correlation ratio thus determine respectively the intensity and spatial organization of heterogeneities in the depth averaged velocity field, therefore controlling the advective backbone for thermal transport inside the fracture.
%


\section{Heat Transfer in Fracture–Matrix Systems}\label{sec:transport}

\subsection{Formulation of the heat transport problem in a fracture-matrix system}

Heat transport in fractured media involves advective and thermal diffusive processes within the mobile fluid phase confined to the fractures, and purely conductive transport within the surrounding impermeable rock matrix. At the fracture scale, the strong geometric confinement justifies a depth-averaged formulation of the energy equation \citep{Bear1972,Berkowitz2002}: transverse thermal gradients equilibrate much faster than longitudinal variations, even in rough-walled fractures \citep{Neuville2013,Ma2019}. Because the aperture is orders of magnitude smaller than the in-plane scales, cross-aperture conduction brings the fluid and the walls to a rapid quasi-equilibrium. The aperture-averaged temperature therefore provides an accurate representation of the thermal field consistent with lubrication–scale analyses in rough fractures and with homogenized descriptions of fluid–solid systems \citep[e.g.][]{Bataille2006,Neuville2010}. Under these conditions, the aperture-averaged temperature satisfies the two-dimensional advection–dispersion equation
\begin{equation}\label{ADE_eq}
\frac{\partial T_{\mathrm{f}}}{\partial t}
+ \mathbf{u}\!\cdot\!\nabla T_{\mathrm{f}}
= D_{\mathrm{f}} \nabla^2 T_{\mathrm{f}}
- \mathcal{B}, 
\qquad \mathbf{x}\in\Omega_{\mathrm{f}},
\end{equation}
where $\mathcal{B}$ is the sink term accounting for heat loss to the matrix, and $D_{\mathrm{f}}=k_{\mathrm{w}}/(\rho_{\mathrm{w}}c_{p,\mathrm{w}})$ is the fluid thermal diffusivity, with $k_{\mathrm{w}}$ and $c_{p,\mathrm{w}}$ denoting the fluid thermal conductivity and specific heat capacity, respectively.

Field observations indicate that fracture spacings in crystalline rocks and geothermal reservoirs often range from metres to tens of metres \citep{Odling1997,Berkowitz2002,Neuman2005,Genter2010}. Smaller spacings also occur in highly fractured fault-damage zones, where fracture density is markedly higher \citep{Brixel2020,Dutler2019}. Over the short and intermediate timescales relevant for fracture–scale heat exchange, the associated thermal penetration depth remains much smaller than these distances, so the rock can be treated as a semi-infinite medium \citep{Neuville2010,Klepikova2021}. At longer times, the penetration depth approaches the matrix-block thickness or the distance to neighbouring fractures, at which point finite-size effects emerge. Other configurations for which this demi-infinite assumption breaks up are those for which most fractures intersect other fractures, as is often the case in crystalline rocks such as granite \citep{bonnetRevGeophys2001}. These conditions fall outside the scope of the present single-fracture analysis.

Within the rock, heat transport is assumed to be purely conductive and predominantly normal to the fracture plane \citep{CarslawJaeger1959}. This approximation is supported by the analysis of \citet{Jung2012}, who showed that, under typical geothermal and crystalline-rock conditions, in-plane diffusion has only a minor influence on the overall heat-exchange dynamics. Under such conditions, heat transfer in the matrix is governed by the one-dimensional semi-infinite diffusion equation:
\begin{equation}\label{heat_eq}
\frac{\partial T_{\mathrm{m}}}{\partial t}
= \alpha_{\mathrm{m}}
\frac{\partial^2 T_{\mathrm{m}}}{\partial {x_3^\prime}^2},
\qquad x_3^\prime\ge0,
\end{equation}
where $\alpha_{\mathrm{m}}=k_\mathrm{r}/(\rho_\mathrm{r}c_{p,\mathrm{r}})$ is the thermal diffusivity of the rock. 

The governing equations \eqref{ADE_eq} and \eqref{heat_eq} are closed by the following initial and boundary conditions.
Initially, the temperature is uniform in both domains, which are at thermal equilibrium with each other:
\begin{equation}
T_{\mathrm{f}}(\mathbf{x},0)=T_0, 
\qquad 
T_{\mathrm{m}}(x_3^\prime,0)=T_0 .
\end{equation}
At times $t>0$ a permanent injection temperature is prescribed at the fracture inlet:
\begin{equation}
T_{\mathrm{f}}(0,x_2,t)=T_1 .
\end{equation}
Periodic conditions are imposed along the lateral boundaries:
\begin{equation}
T_{\mathrm{f}}(x_1,0,t)=T_{\mathrm{f}}(x_1,L,t), \qquad
\frac{\partial T_{\mathrm{f}}}{\partial x_2}(x_1,0,t)
=
\frac{\partial T_{\mathrm{f}}}{\partial x_2}(x_1,L,t).
\end{equation}
The rock matrix is modeled, for each wall, as a semi–infinite half–space normal to the fracture plane, with the following far–field condition:
\begin{equation}
T_{\mathrm{m}}(\mathbf{x},\,x_3^\prime\to \infty,\,t)=T_0 .
\end{equation}
Finally, temperature is assumed to be continuous across the fluid--rock interface on both walls:
\begin{equation} \label{eq:temp_cont}
T_{\mathrm{m}}(\mathbf{x},0,t)=T_{\mathrm{f}}(\mathbf{x},t).
\end{equation}
Because the fracture aperture is much smaller than the in--plane characteristic length scales, heat conduction across the aperture rapidly equilibrates the fluid and wall temperatures on timescales that are short compared to those that are characteristic of longitudinal transport. Consequently, the depth-average temperature field inside the fracture can be considered equal to what the temperature would be in the vicinity of the solid wall and at the same in-plane position if a 3D description of the fracture 
were considered. Hence continuity between the depth-average fluid temperature and the temperature in the rock matrix can be assumed at the rock-matrix interface. This assumption has been widely adopted in laboratory, numerical, and theoretical studies of
fracture--matrix heat exchange in both smooth and rough--walled fractures
\citep{Vinsome1980,Neuville2013,Gisladottir2016,Cherubini2017,Klepikova2021}.

Solving the one–dimensional diffusion problem \eqref{heat_eq} in the semi–infinite matrix with the time–varying boundary temperature \eqref{eq:temp_cont}, and applying Duhamel’s principle, yields the matrix temperature anomaly
\begin{equation}\label{Tm_erfc_dimless}
\Delta T_{\mathrm{m}}(\mathbf{x},x_3^\prime,t)
= \int_0^t
\frac{\partial \Delta T_{\mathrm{f}}(\mathbf{x},\tau)}{\partial \tau}\,
\operatorname{erfc}\!\left(
\frac{x_3^\prime}{2\sqrt{\alpha_{\mathrm{m}}(t-\tau)}}\right)\,d\tau,
\qquad
\Delta T_{\mathrm{f}}(\mathbf{x},t)
= \frac{T_{\mathrm{f}}(\mathbf{x},t)-T_0}{T_1-T_0}.
\end{equation}
where $\operatorname{erfc}(\cdot)$ is the complementary error function.

Differentiating with respect to \(x_3^\prime\) and invoking Fourier’s law gives the interfacial heat flux
\begin{equation}\label{flux_q_dim}
q_{\mathrm{m}}(\mathbf{x},t)
=
\frac{2\,\phi_{\mathrm{m}}\,k_{\mathrm{r}}(T_1-T_0)}{\sqrt{\pi\,\alpha_{\mathrm{m}}}}
\frac{d}{dt}
\int_0^t
\frac{\Delta T_{\mathrm{f}}(\mathbf{x},\tau)}{\sqrt{t-\tau}}\,d\tau,
\end{equation}
where
\begin{equation}\label{eq:phim}
\phi_{\mathrm{m}}
= \phi_\mathrm{r}
+ (1-\phi_\mathrm{r})
\frac{\rho_{\mathrm{r}}c_{p,\mathrm{r}}}
     {\rho_{\mathrm{w}}c_{p,\mathrm{w}}}
\end{equation}
is the dimensionless volumetric heat–capacity ratio of the saturated matrix, obtained by normalizing its volumetric heat capacity to that of the fracture fluid. Here, $\phi_\mathrm{r}$ denotes the rock porosity, which enters as the weighting factor used to construct the matrix volumetric heat capacity and reflects the relative contributions of fluid and solid to thermal storage \citep{ruiz-2014-analytical,Gisladottir2016}.

Equation~\eqref{flux_q_dim} expresses the wall flux as a Volterra convolution with kernel \((t-\tau)^{-1/2}\), encoding the semi–infinite diffusive memory of the matrix.

This yields the fracture–scale sink term to be considered in Eq.~\ref{ADE_eq}
\begin{equation}\label{B_def}
\mathcal{B}(\mathbf{x},t)
= \frac{q_{\mathrm{m}}(\mathbf{x},t)}
       {\rho_{\mathrm{w}}c_{p,\mathrm{w}}\,a(\mathbf{x})}
= \frac{2\,\phi_{\mathrm{m}}\,k_{\mathrm{r}}(T_1-T_0)}
       {\rho_{\mathrm{w}}c_{p,\mathrm{w}}\,a(\mathbf{x})\sqrt{\pi\,\alpha_{\mathrm{m}}}}
\frac{d}{dt}
\int_0^t
\frac{\Delta T_{\mathrm{f}}(\mathbf{x},\tau)}{\sqrt{t-\tau}}\,d\tau.
\end{equation}
Equations~\eqref{flux_q_dim}–\eqref{B_def} thus recover the fractional \((t-\tau)^{-1/2}\) kernel associated with diffusion into a semi–infinite solid. Note that Equation~\ref{ADE_eq} associated to the sink term~\ref{B_def} remains linear, ensuring that the memory kernel, which represents the probability distribution of residence
times associated with conductive exchange in the matrix, is independent of the
sign of $T_1-T_0$, and therefore equally applies to fluid heating and
cooling scenarios.

In geothermal studies, the relative importance of advective versus conductive heat transfer is typically assessed using the thermal Péclet number, which quantifies the ratio between the advective heat flux ($q_{\mathrm{adv}}$) transported by the flowing fluid and the conductive heat flux ($q_{\mathrm{cond}}$) through the surrounding rock matrix \citep{Gossler2019,Klepikova2021}:
\begin{equation}\label{eq:Peclet}
\Pen = \frac{q_{\mathrm{adv}}}{q_{\mathrm{cond}}} = \frac{\rho_\mathrm{w} c_{p,\mathrm{w}}\, U_\mathrm{c}\, (T_1 - T_0)}{k_\mathrm{r} \, (T_1 - T_0)/a_\mathrm{m}} = \frac{\rho_\mathrm{w} c_{p,\mathrm{w}}\, U_\mathrm{c}\, a_\mathrm{m}}{k_\mathrm{r}}.
\end{equation}
The same temperature difference appears in both $q_{\mathrm{adv}}$ and $q_{\mathrm{cond}}$, which reflects the fact that, at initial times, the temperature difference between the fracture matrix interface in contact with the injected fluid and the rock away from that interface is $T_1-T_0$, equal to the temperature difference between the injected fluid and the resident fluid. This is because the temperature in the fluid can be considered uniform across the aperture, for two reasons: (i) the typical cross-aperture heat conduction time is several orders of magnitudes smaller than the typical advection time along the fracture length, so cross-aperture thermal gradients equilibrate rapidly by diffusion across the aperture, under Stokes flow; (ii) in applications such as geothermy, the injected fluid occupies the entire fracture aperture at the injection point, and has a uniform temperature, so the out-of-plane uniformity of the injected fluid's temperature is ensured from the start of the injection. This formulation of the thermal Péclet number is consistent with classical heat-transfer models in fractured rocks \citep{Marsily1993,Ge1998}, where heat exchange is controlled by fluid advection inside the fracture and conduction into the surrounding rock.

It is worth noting that temperature-dependent fluid properties may influence the transient dynamics but do not alter the fundamental structure of the nonlocal matrix-diffusion kernel. Large temperature contrasts, such as those occurring during cold-water injection in enhanced geothermal systems, can temporarily reduce water viscosity by up to an order of magnitude \citep{Wagner2002}, slightly modifying the local advective velocity. These variations decay rapidly as thermal equilibration develops and effectively rescale the characteristic flow speed without affecting the heavy-tailed nature of matrix trapping. Thermal conductivities and heat capacities of water and rock vary only weakly with temperature, so the constant-property approximation remains suitable for the fracture-scale processes analysed here \citep{Okoroafor2022}.

\subsection{Stochastic representation of heat transport via time-domain random walk}

We do not explicitly solve Eq.~\ref{ADE_eq} with sink term~\ref{B_def}, but use a time-domain random walk (TDRW) formulation which is asymptotically equivalent, in the limit of a sufficiently large number of particles, to solving the Eulerian heat transport equation. This formulation is thus also equally suited to scenarios where the injected water is warmer or cooler than the rock.

The finite-volume discretization of equation~\eqref{ADE_eq}, following \citet{Delay2002}, reads:
\begin{equation}\label{discrete_ADE}
V_i\,\frac{\partial T_{\mathrm{f},i}(t)}{\partial t}
= \sum_{j\in\sigma(i)} b_{ij}\,V_j\,T_{\mathrm{f},j}(t)
- \sum_{j\in\sigma(i)} b_{ji}\,V_i\,T_{\mathrm{f},i}(t)
- V_i\,\mathcal{B}_i(t),
\end{equation}
where $V_i$ denotes the volume of the fracture control volume, $T_{\mathrm{f},i}(t)$ is the temperature associated with cell $i$, and $j\in\sigma(i)$ denotes the set of its four nearest neighbours $\sigma(i)=\{N,S,E,W\}$. The exchange coefficients $b_{ij}$ are defined as
\begin{equation}\label{discrete_ADE_coeff}
b_{ij}=\frac{S_{ij}\,D_\mathrm{f}}{V_j\,\Delta x}
+\frac{S_{ij}\,|u_{ij}|}{2V_j}
\left(\frac{u_{ij}}{|u_{ij}|}+1\right),
\end{equation}
where $S_{ij}$ is the surface area of the shared interface
between cells $i$ and $j$, and $u_{ij}>0$ denotes a velocity directed from $j$ to $i$
(negative otherwise).

The last term in Eq.~\eqref{discrete_ADE}, $\mathcal{B}_i(t)$, accounts for the local
heat exchange between the fluid and the surrounding rock matrix, as defined in Eq.~\eqref{B_def}.
Using the dimensionless temperature anomaly 
$\Delta T = (T-T_0)/(T_1-T_0)$, it reads
\begin{equation}\label{discrete_Bi}
\mathcal{B}_i(t)=\frac{2\,\phi_\mathrm{m}\,k_\mathrm{r}\,(T_1-T_0)}   {\rho_\mathrm{w}c_{p,\mathrm{w}}\,a_i\,\sqrt{\pi\,\alpha_\mathrm{m}}}\,\frac{d}{dt}\int_0^t\frac{\Delta T_{\mathrm{f},i}(\tau)}{\sqrt{t-\tau}}\,d\tau,
\end{equation}
which represents the temporally nonlocal (memory) response of the rock matrix and
corresponds to a fractional-order sink term with kernel $(t-\tau)^{-1/2}$.

This term acts as a distributed sink in the fracture energy balance,
and ensures continuity of temperature and heat flux at the fracture–matrix interface.
Its inclusion in Eq.~\eqref{discrete_ADE} preserves energy conservation in the discrete system
and provides the physical basis for the transition to the Master Equation
and the time-domain random walk (TDRW) formulation described below.

Equation~\eqref{discrete_ADE} expresses the conservative balance of
thermal energy within each finite volume and can be recast as a Master
Equation by introducing the conserved energy variable
$g_i(t)=V_i\,T_{\mathrm{f},i}(t)$, which represents both the mobile
thermal energy stored in cell $i$ and the corresponding density of TDRW 
particles used to sample its evolution. 

In the TDRW formulation, particles act as Lagrangian tracers that sample the
stochastic residence--time structure of the coupled fracture--matrix system. 
Thermal transport is represented through the ensemble evolution of the
conserved thermal energy $g_i(t)$ carried by each particle, which follows the
same balance as the finite--volume advection--diffusion equation with a
fracture--matrix sink term accounting for conductive heat exchange with the
matrix. The corresponding aperture--averaged fracture temperature then follows
as $T_{\mathrm{f},i}(t)=g_i(t)/V_i$.

This reinterpretation enables a stochastic description of particle dynamics, in which the temporal evolution of $g_i(t)$ is governed by transition probabilities between neighboring cells. Specifically, the probability $\mathbb{P}_{ij}$ of a particle jumping from cell $j$ to a neighbouring cell $i$, and the associated mobile residence time $\tau_j$, are given by
\begin{equation}\label{eq:transit_prob_time}
\mathbb{P}_{ij} = \mathbb{P}(\mathbf{x}_i \mid \mathbf{x}_j) = \frac{b_{ij}}{\sum\limits_{k \in \sigma(j)} b_{kj}}, \quad \mathrm{and} \quad \tau_j=\frac{1}{\sum_{k\in\sigma(j)}b_{kj}}.
\end{equation}
Consequently, the evolution of the particle density is described by the following Master Equation:
\begin{equation}\label{eq:master_eq}
\frac{d g_i(t)}{d t} = \sum_{j \in \sigma(i)} \mathbb{P}_{ij} \frac{g_j(t)}{\tau_j} - \frac{g_i(t)}{\tau_i}\,-V_i\,\mathcal{B}_i(t),
\end{equation}
where the last term represents the continuous loss of mobile energy due to conductive trapping within the matrix. 

In the lattice-based random walk, the particle located at $\mathbf{x}^{(n)}=\mathbf{x}_j$ 
selects one of the neighbouring cells $i\in\sigma(j)$ with probability $\mathbb{P}_{ij}$.  
Denoting by $\boldsymbol{\xi}_{ij}$ the displacement from the centre of cell $j$ to the centre 
of cell $i$, the update reads as
\begin{equation}\label{eq:recursive_eq}
\mathbf{x}^{(n+1)} = \mathbf{x}^{(n)} + \boldsymbol{\xi}_{ij},
\qquad
t^{(n+1)} = t^{(n)} + \theta_{\mathrm{f},j}^{(n)},
\end{equation}
where the mobile residence time $\theta_{\mathrm{f},j}^{(n)}$ associated with the $n$-th jump is drawn from the exponential probability density function
\begin{equation}\label{eq:exp_transition_time}
\psi_{\mathrm{f},j}(t)
=\frac{1}{\tau_j}\exp\!\left(-\frac{t}{\tau_j}\right).
\end{equation}
Equivalently, $\theta_{\mathrm{f},j}^{(n)}$ is sampled as $\theta_{\mathrm{f},j}^{(n)} = -\tau_j \ln(\eta_1)$ with $\eta_1\sim\mathcal{U}(0,1]$. This recursive propagation scheme mirrors that of CTRW-based models, where particle trajectories evolve through a sequence of spatial jumps and exponentially distributed mobile residence times determined by local finite-volume exchange coefficients \citep{Dentz2004}. 

\subsection{Matrix conduction and trapping-time formulation}\label{sec:matrix_conduction}
Matrix conduction is incorporated using the semi-analytical framework of \citet{Painter2005}, which relates the trapping time in a semi-infinite matrix to the fracture residence time via:
\begin{equation}\label{eq:diffusion_time}
\theta_{\mathrm{m},j}=\bigg[\frac{\phi_\mathrm{m}\sqrt{\alpha_\mathrm{m}}\theta_{\mathrm{f},j}}{a_j\: \operatorname{erfc}^{-1}(\eta_2)}\bigg]^2,
\end{equation}
where $\eta_2 \in \mathcal{U}(0,1]$ is a uniformly distributed random variable. The parameter $\phi_\mathrm{m}$ naturally emerges in the stochastic formulation as a scaling factor ensuring energy conservation between the mobile and immobile domains \citep{Haggerty1995,Delay2001,Painter2005}.

The stochastic formulation of Eq.~\eqref{eq:diffusion_time} provides a discrete counterpart to the continuous source term $\mathcal{B}$ introduced in Eq.~\eqref{B_def}.
While $\mathcal{B}$ represents the nonlocal energy exchange between the fracture and the surrounding matrix in the continuum (ADE) formulation, the random variable $\theta_{\mathrm{m},j}$ embodies the same process at the particle level.
Each trapping event corresponds to a transient conductive exchange across the fracture–matrix interface; ensemble averaging of these events recovers the nonlocal kernel in Equation~\eqref{B_def}.

Equation~\eqref{eq:diffusion_time} implies that the matrix trapping time $\theta_{\mathrm{m},j}$ follows a Lévy–Smirnov distribution with advective conditioning (see Appendix \ref{appA}):
\begin{equation}\label{eq:trapping_time_dist}
\psi_{\mathrm{m},j}\!\left(t \,\big|\, \theta_{\mathrm{f},j}\right)
= \frac{\phi_{\mathrm{m}}\,\sqrt{\alpha_{\mathrm{m}}}\,\theta_{\mathrm{f},j}}
       {a_j\,\sqrt{\pi}}\,
  t^{-3/2}\,
  \exp\!\left[
    -\,\frac{\phi_{\mathrm{m}}^{2}\,\alpha_{\mathrm{m}}\,\theta_{\mathrm{f},j}^{2}}
           {a_j^{2}\,t}
  \right],
  \qquad t>0.
\end{equation}
The Lévy--Smirnov distribution arises here in a conditioned form: the characteristic scale of the matrix trapping time depends explicitly on the realised advective residence time in the fracture. Physically, this reflects the fact that conductive heat exchange with the matrix can only occur during the time a fluid parcel remains in contact with the fracture walls. Advective transport therefore sets the time window over which diffusive exchange is active, while matrix conduction provides the intrinsic non-local memory of the system.

This Lévy--Smirnov distribution is the first--passage--time law for a one--dimensional Brownian particle to reach an absorbing boundary. It has a heavy tail, $\psi_{\mathrm{m},j}(t)\propto t^{-3/2}$, and an infinite mean, implying a non-negligible probability of very long trapping events. In the present context, Eq.~\eqref{eq:diffusion_time} represents the random time required for heat to diffuse from the fracture--matrix interface into an effectively semi-infinite rock half-space. The $t^{-3/2}$ tail is therefore a direct consequence of Fickian diffusion in the matrix and of the semi-infinite geometry. Its impact is to imprint a enduring memory on the in-fracture dynamics, yielding non-Fickian signatures (e.g. power-law late-time behaviour) in fracture-scale observables such as breakthrough curves and thermal power.

This framework yields a L\'evy--Smirnov trapping-time distribution that follows directly from the diffusion first-passage problem in a semi-infinite matrix. As a result, the heavy-tailed behaviour is physically derived rather than prescribed, and no distribution of exchange rates is required. In contrast, classical multirate mass-transfer (MRMT) models approximate matrix exchange as a superposition of first-order processes with finite mean waiting times. Here, a single, heavy-tailed kernel captures the broad range of matrix residence times generated by pore-scale diffusion, exhibiting the characteristic $t^{-1/2}$ and $t^{-3/2}$ regimes associated with semi-infinite diffusion.

To incorporate matrix diffusion into the random walk, the particle’s 
transition time~$\theta_j$ in voxel~$j$ is written as the sum of a mobile 
advective period and a matrix--diffusion trapping interval,
\begin{equation}\label{Recursive_EQ}
t^{(n+1)}
= t^{(n)} + \theta_j^{(n)}
= t^{(n)} + \theta_{\mathrm{f},j}^{(n)} + \theta_{\mathrm{m},j}^{(n)} .
\end{equation}
Accordingly, the total residence time in cell~$j$ consists of a mobile 
advective duration~$\theta_{\mathrm{f},j}$ followed by a diffusive trapping 
time~$\theta_{\mathrm{m},j}$. This residence--time distribution stems from the 
exact statistical mixture of the exponential mobile--time law with the 
conditional L\'evy--Smirnov first--passage distribution governing diffusion 
into the surrounding matrix.

The resulting Laplace transform of the total residence-time density is
\begin{equation}
\tilde\psi_j(s)
=
\frac{1}{1+\tau_j s + A_j\sqrt{s}},
\qquad
A_j = \frac{2\,\phi_{\mathrm{m}}\sqrt{\alpha_{\mathrm{m}}}\,\tau_j}{a_j},
\label{eq:psi_laplace}
\end{equation}
The corresponding memory kernel follows from the standard CTRW relation
\begin{equation}
\tilde{\mathcal K}_j(s)
=
\frac{s\,\tilde\psi_j(s)}{1-\tilde\psi_j(s)}
=
\frac{\sqrt{s}}{A_j + \tau_j\sqrt{s}},
\label{eq:K_laplace}
\end{equation}
whose time-domain expression is
\begin{equation}
\mathcal{K}_j(t)
=
\tau_j^{-1}\,\delta(t)
-
\frac{A_j}{\tau_j^{2}\sqrt{\pi t}}
+
\frac{A_j^{2}}{\tau_j^{3}}
\exp\!\Bigg(\frac{A_j^2}{\tau_j^2} t\Bigg)
\operatorname{erfc}\!\Bigg(\frac{A_j}{\tau_j}\sqrt{t}\Bigg),
\qquad t>0.
\label{eq:K_time}
\end{equation}
The kernel $\mathcal{K}_j(t)$ features the characteristic short-time singularity $\mathcal{K}_j(t)\sim t^{-1/2}$ (for the non-impulsive part) and a long-time algebraic tail $\mathcal{K}_j(t)\sim t^{-3/2}$, consistent with diffusive exchange with a semi-infinite matrix. These asymptotics coincide with the nonlocal transport kernels of \citet{Haggerty1995}, \citet{Dentz2004}, and \citet{Painter2005}, confirming that the present TDRW formulation is an exact stochastic analogue of the corresponding nonlocal advection--diffusion equation for fracture--matrix heat exchange. The complete derivation of the kernel, including its asymptotic behaviour at short and long times, is provided in Appendix~\ref{appB}.

In this work, we analyze heat transfer in a fracture under the scenario of continuous injection of hot fluid at the inlet. The same formulation applies to cold-fluid injection, as only temperature differences enter the governing equations. Because explicitly simulating a continuous injection is computationally prohibitive, we instead simulate particle dynamics for a single “slug” (i.e., instantaneous) injection, which substantially reduces the number of particles required. We then leverage the fact that the cumulative distribution function ($CDF$) of particle arrival times at any location for an instantaneous injection is mathematically equivalent to the solution for continuous injection, allowing us to reconstruct continuous‑injection BTCs from instantaneous injection random walks. 

The probability density function (PDF) of the arrival time at node $\mathbf{x}_j$ is estimated from the series of transition times $t_i=\{t_1, t_2, \dots, t_{N(\mathbf{x}_j)}\}$ of the $N(\mathbf{x}_j)$  particles that have passed through that node, as:
\begin{equation}
\label{eq:PDF_arrival_times}
f(\mathbf{x}_j,t)
=
\frac{1}{N(\mathbf{x}_j)}
\sum_{i=1}^{N(\mathbf{x}_j)}
\delta\!\bigl(t - t_i(\mathbf{x}_j)\bigr),
\end{equation}
Subsequently, the thermal anomaly in the fracture for the continuous injection boundary condition, 
\begin{equation}
\label{eqthermal_anomaly_fracture}
\Delta T_{\mathrm{f}}(\mathbf{x}, t) = \frac{T_\mathrm{f}(\mathbf{x}, t)-T_0}{T_1-T_0},
\end{equation}
can be indirectly derived by evaluating the $CDF$ of the arrival times for a given time $t$ at each node, as:
\begin{equation}
\label{eq:CDF_arrival_times}
\Delta T_{\mathrm{f}}(\mathbf{x}_j,t)
= F(\mathbf{x}_j,t)
= \frac{1}{N(\mathbf{x}_j)}
  \sum_{i=1}^{N(\mathbf{x}_j)}
  \mathbb{I}\!\left( t_i(\mathbf{x}_j) \le t \right).
\end{equation}
where $\mathbb{I}$ is an indicator function that equals 1 if the condition inside the braces is true and 0 otherwise.

Indeed, $\Delta T_{\mathrm{f}}(\mathbf{x}_j, t)$ is equivalent to the cumulative distribution function ($CDF$) $F(\mathbf{x}_j, t)$ of the arrival times of all particles passing through the location $\mathbf{x}_j$. The $CDF$ is numerically computed by summing the PDF values for all times $k$ less than or equal to $t$, which is equivalent to calculating the proportion of particles that arrived at node $\mathbf{x}_j$ by time $t$. This on-the-fly approach allows reducing the memory usage considerably.

\subsection{Example of temperature anomaly evolution in time}

\begin{figure}
\centerline{\includegraphics[width=0.7\textwidth]{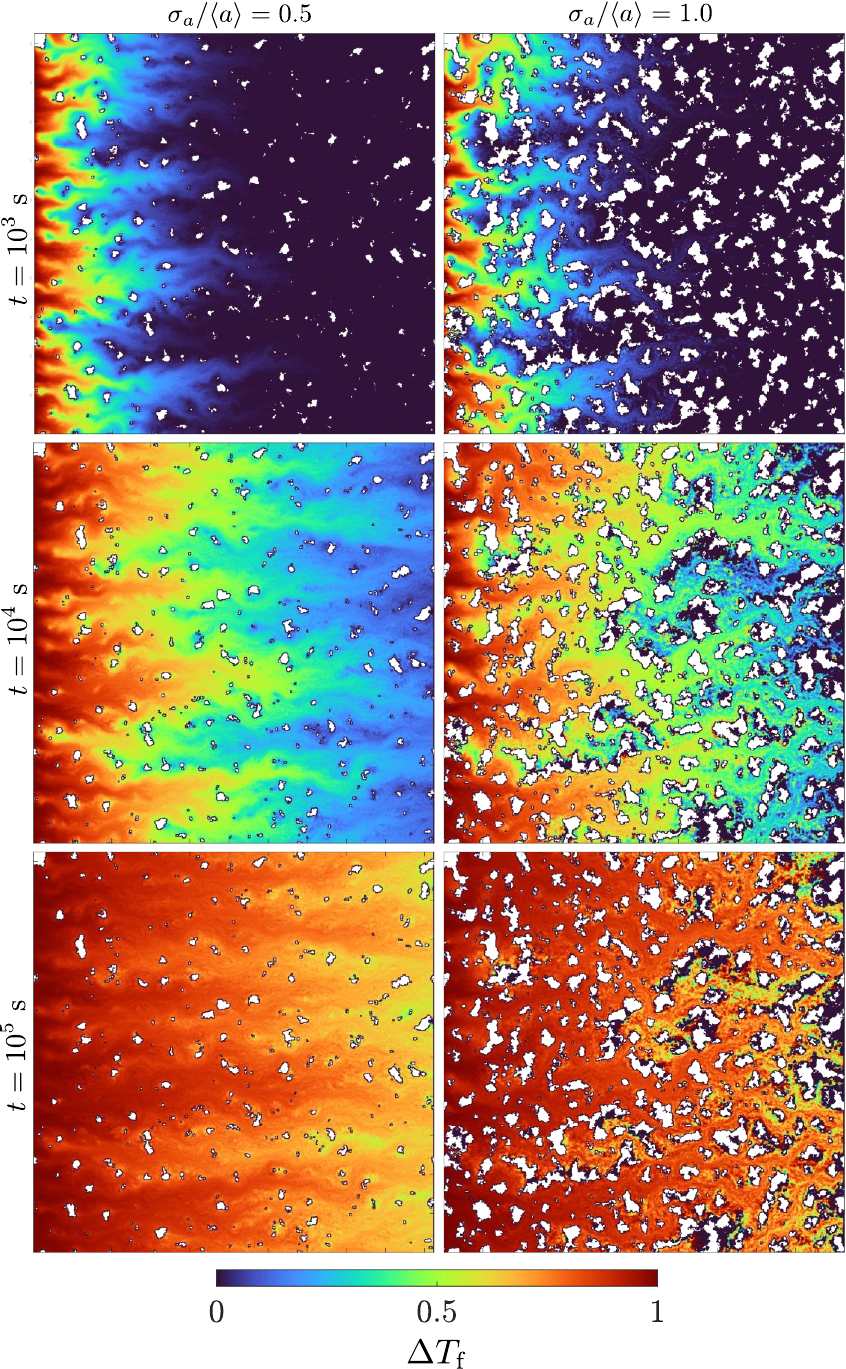}}
\caption{Normalized temperature anomaly fields $\Delta T_{\mathrm{f}}$ for two
levels of aperture heterogeneity, $\sigma_a/\langle a\rangle = 0.5$ (left) and
$\sigma_a/\langle a\rangle = 1.0$ (right), for a fracture correlation length
$L_\mathrm{c} = 0.1~\mathrm{m}$ ($L/L_\mathrm{c}= 2^5$) and $\Pen=50$. Rows correspond to times $t = 10^3$, $10^4$, and
$10^5~\mathrm{s}$ (top to bottom). Increasing heterogeneity enhances flow
localization and enlarges contact regions (white), yielding stronger spatial
variability and more persistent thermal contrasts.}
\label{Fig2}
\end{figure}
Figure~\ref{Fig2} shows the evolution of the normalized temperature anomaly 
$\Delta T_{\mathrm{f}}$ for two levels of aperture heterogeneity,
$\sigma_a/\langle a\rangle = 0.5$ (left column) and
$\sigma_a/\langle a\rangle = 1.0$ (right column), evaluated for $\Pen=50$. Each row corresponds to a
different time, $t = 10^3$, $10^4$, and $10^5~\mathrm{s}$ (from top to bottom),
illustrating how heat transport responds to increasing structural disorder. Heat is
injected uniformly along the left boundary and transported downstream through the
variable-aperture field, producing markedly different temperature patterns
depending on the degree of heterogeneity.

At early times ($t = 10^3~\mathrm{s}$), the thermal front advances preferentially
through high-aperture, high-transmissivity channels while remaining strongly
retarded near contact zones, generating a distinctly fingered structure. As time
progresses ($t = 10^4$–$10^5~\mathrm{s}$), conduction within the fracture and across
the fracture–matrix interface gradually smooths the field, yet significant thermal
contrasts persist owing to the strong aperture variability.

Increasing aperture variability enhances flow localization and small-scale
tortuosity, as evident from the comparison between the two figure columns. For the
moderately heterogeneous case ($\sigma_a/\langle a\rangle = 0.5$), the velocity
field remains relatively smooth and heat is transported through broad,
interconnected pathways, leading to a quasi-planar front and progressive spatial
homogenization of $\Delta T_{\mathrm{f}}$. In contrast, when
$\sigma_a/\langle a\rangle = 1.0$, the thermal anomaly becomes highly intermittent:
heat is confined to narrow, filamentary conduits separated by extensive
low-velocity regions that remain cold for long periods. These quasi-stagnant zones
act as local thermal traps, storing heat that is later released via conduction,
thereby sustaining pronounced temperature contrasts even at late times.

\subsection{Outlet breakthrough curve and arrival-time statistics}
The breakthrough behaviour at the fracture outlet provides a complementary and more detailed description of heat-transport dynamics than spatial moments alone.
Although particle trajectories are generated using an impulse (slug) injection, the physical quantity of interest is the breakthrough response under continuous injection.
In the stochastic formulation, the continuous-injection breakthrough curve $BTC$ is obtained directly from the cumulative distribution of arrival times associated with the impulse injection, exploiting the equivalence between these two representations.
In this sense, $BTC$ describes the temporal release of heat at the fracture outlet.

Specifically, the fraction of particles that have reached the outlet by time $t$ is quantified by the cumulative distribution function,
\begin{equation}\label{eq:BTC_def}
BTC(t)=CDF(t)=\mathbb{P}\{\tau \le t\},
\end{equation}
where $\tau$ denotes the travel time of an individual particle from injection to detection.
To emphasize late-time behaviour, we also consider the complementary cumulative distribution function,
\begin{equation}\label{eq:CCDF}
CCDF(t)=1-BTC(t)=\mathbb{P}\{\tau > t\},
\end{equation}
The $CCDF$ form emphasizes late-time behavior and highlights the persistence of slow processes such as conductive exchange with the rock matrix and trapping in low-velocity regions, which may not be fully captured by lower-order moments of the displacement distribution. For an instantaneous release of particles, the $CCDF$ represents the fraction of particles that have not yet reached the outlet. It serves as a statistical fingerprint of the interplay between advection, dispersion, and conductive exchange, allowing us to complement the analysis of mean displacement and variance by revealing the non-Gaussian, heavy-tailed characteristics of thermal transport and by directly linking late-time behavior to the underlying memory effects associated with diffusive trapping.

In the ideal advection-dominated limit, where the velocity field is uniform and conductive exchange is negligible, the arrival-time distribution is narrow and the $CCDF$ decays sharply. In contrast, the presence of velocity contrasts, fracture heterogeneity, and matrix conduction induces a broad distribution of travel times, leading to a gradual decay of $CCDF$ and the emergence of heavy tails. Such tailing behavior, widely reported in solute and heat transport through heterogeneous fractures, reflects the combined influence of advective variability and diffusive exchange across fracture–matrix interfaces.

During the Monte Carlo simulations, the $CCDF$ associated with the breakthrough curve is computed on the fly, without explicitly constructing a probability density function. Each time a particle reaches the control plane, its arrival time~$\tau_i$ is recorded and assigned to a discrete time bin~$t_k$. The cumulative number of arrivals up to~$t_k$ provides the empirical estimator of the cumulative distribution function ($CDF$):
\begin{equation}\label{eq:CDF_empirical}
CDF(t_k) = \frac{1}{N_\mathrm{p}} 
\sum_{i=1}^{N_\mathrm{p}} \mathbb{I}\bigl[\tau_i \le t_k\bigr].
\end{equation}
The complementary cumulative distribution function, corresponding to the breakthrough curve, is then obtained as
\begin{equation}\label{eq:CCDF_empirical}
CCDF(t_k) = \frac{N_\mathrm{p} - N_{\le t_k}}{N_\mathrm{p}}.
\end{equation}
where $N_{\le t_k}$ denotes the cumulative number of particles whose arrival time at the outlet control plane satisfies $\tau_i \le t_k$.

This formulation enables an incremental update of $CCDF$ during the simulation, with negligible memory cost. Particles that do not reach the outlet within the maximum simulation time~$t_\mathrm{max}$ are treated as right-censored events, so that $CCDF(t_\mathrm{max})$ directly quantifies the fraction of heat still retained within the fracture–matrix system.
In the thermal formulation, the breakthrough curve can be equivalently expressed as the cross‐sectionally averaged normalized outlet temperature,
\begin{equation}\label{eq:BTC_mean_def}
BTC(t)
= \frac{1}{L}\int_0^{L}
\frac{T_\mathrm{f}(x_1=L,x_2,t)-T_0}{T_1-T_0}\,dx_2,
\end{equation}
This definition links the stochastic and thermal formulations, allowing $BTC$ to be interpreted as the cumulative distribution of particle arrival times, while its complement $CCDF$ represents the fraction of heat still retained within the fracture--matrix system. Equivalently, $BTC$ corresponds to the normalized mean temperature anomaly at the outlet. These interpretations are fully consistent, since particles that have not yet exited the fracture directly account for the heat still stored within the fracture--matrix system at time~$t$.

\section{Model Validation}~\label{sec:validation}
\begin{figure}
\centerline{\includegraphics[width=\textwidth]{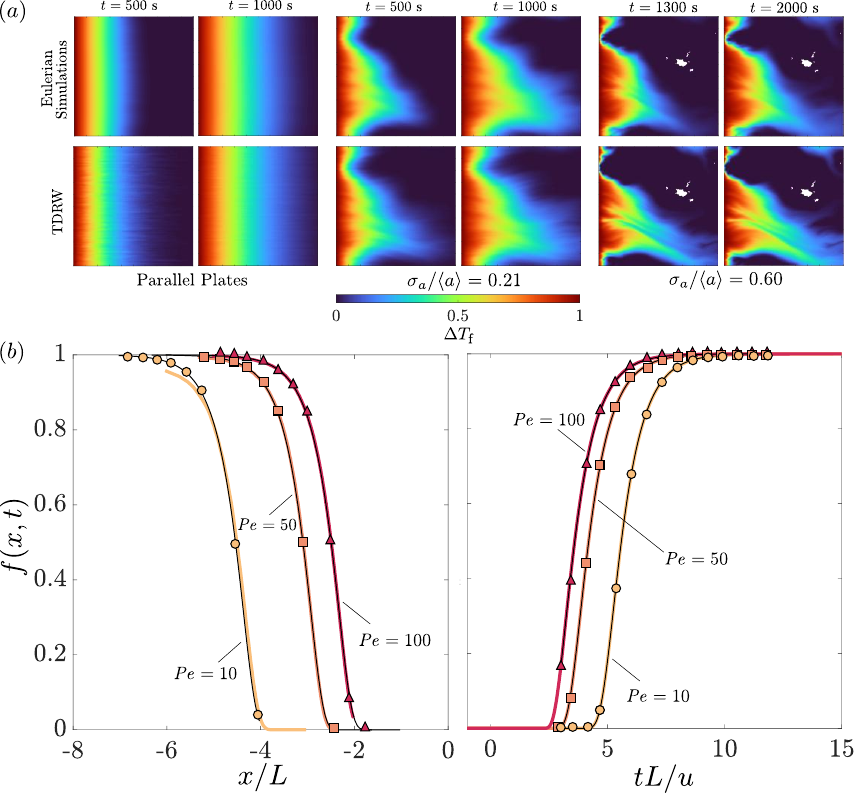}}
\caption{(a) Comparison between the results from a resolution of the Eulerian equations (COMSOL Multiphysics\textsuperscript{\textregistered})  
(top row) and our TDRW predictions (bottom row) for three aperture configurations. 
Columns correspond to: (i) a parallel–plate fracture with mechanical aperture 
$a_{\mathrm{m}} = 0.001~\mathrm{m}$; (ii) a moderately heterogeneous aperture field with 
closure ratio $\sigma_a/\langle a\rangle = 0.21$; and (iii) a strongly heterogeneous field with 
$\sigma_a/\langle a\rangle = 0.60$, for which the mechanical aperture is set to 
$a_{\mathrm{m}} = 3.3\times 10^{-4}~\mathrm{m}$ to maintain comparable wall contact. 
Times are indicated in each panel, and the color map shows the normalized thermal 
anomaly $\Delta T_{\mathrm{f}} = (T_{\mathrm{f}} - T_0)/(T_1 - T_0)$. 
(b) Comparison between our TDRW predictions and the theoretical prediction of \citet{Lauwerier1955} for a parallel plate geometry, for 
$\Pen = \{10,\,50,\,100\}$: spatial temperature profiles (left) and outlet breakthrough curves (right). Physical and numerical parameters are listed in Table~\ref{Tab1}.}
\label{Fig3}
\end{figure}

The numerical scheme adopted in this work has been validated by comparison with
Eulerian simulations performed with (i) the finite-element-based COMSOL
Multiphysics\textsuperscript{\textregistered} software, in which heat conduction
in the rock matrix is treated as a fully three-dimensional diffusion process
(Figure~\ref{Fig3}a), and (ii) the following analytical solution
(Figure~\ref{Fig3}b):
\begin{equation}
\label{eq:lauwerier_profile_anomaly}
\Delta T_{\mathrm{f}}(x,t)
=
\operatorname{erfc}\!\left(
\frac{\phi_{\mathrm{m}}\sqrt{\alpha_{\mathrm{m}}}\,x}
     {a\,u\,\sqrt{t - x/u}}
\right),
\qquad t > x/u .
\end{equation}
which corresponds to the classical solution derived by \citet{Lauwerier1955} for high–Péclet number heat transport in an
infinite parallel-plate fracture of constant aperture embedded in a homogeneous
rock matrix. In this configuration, thermal transport within the fracture is purely advective along the longitudinal direction $x_1$. Heat exchange with the surrounding matrix is entirely due to conduction within the matrix normal to the fracture walls, and the matrix is treated as a semi-infinite medium in that direction. This yields a one-dimensional advective transport problem in the fracture coupled to one-dimensional conductive diffusion in the matrix.

Figure~\ref{Fig3}a presents a direct comparison between the results from our TDRW simulations and those from Eulerian 
COMSOL Multiphysics\textsuperscript{\textregistered} simulations, 
for three aperture configurations of increasing heterogeneity. 
The TDRW method reproduces the spatial structure of the thermal anomaly 
field $\Delta T_{\mathrm{f}}$ predicted by the Eulerian simulations with high fidelity across all cases considered. In the parallel–plate geometry, 
the agreement between the two models is essentially exact. For the moderately 
heterogeneous fracture ($\sigma_a/\langle a\rangle = 0.21$), both COMSOL and TDRW predict 
the same deformation of the thermal front induced by preferential flow channels, 
demonstrating the ability of the stochastic scheme to capture geometrically driven 
advective–conductive patterns. Even in the strongly heterogeneous configuration 
($\sigma_a/\langle a\rangle = 0.60$), characterised by enhanced wall interpenetration and 
extended contact zones, the TDRW approach successfully reproduces the large–scale thermal 
structures observed in the Eulerian reference, with only minor smoothing of small–scale 
gradients due to its particle-based representation.

Figure~\ref{Fig3}(b) further corroborates this agreement by comparing spatial temperature profiles 
and breakthrough curves obtained with our TDRW scheme to those predicted by the \citep{Lauwerier1955} analytical solution, for a parallel plate geometry and for different Péclet numbers. The TDRW results closely follow the 
Lauwerier solution across all $\Pen$.

\begin{table}
\centering
\caption{Comparison between TDRW and COMSOL temperature fields for different fracture closures and observation times. The normalized root-mean-square error (NRMSE) and coefficient of determination ($R^2$) quantify the agreement between the two models.}
\label{Tab2}
\renewcommand{\arraystretch}{1.2}
\begin{tabular}{cccc}
\toprule
 $\sigma_a/\langle a\rangle$ & \textbf{Time [s]} & \textbf{NRMSE} & \textbf{$R^2$} \\
\midrule
Parallel Plates & 500  & 0.026 & 0.992 \\
Parallel Plates & 1000 & 0.026 & 0.990 \\
0.21 & 500  & 0.022 & 0.994 \\
0.21 & 1000 & 0.018 & 0.996 \\  
0.60 & 1300 & 0.094 & 0.887 \\
0.60 & 2000 & 0.098 & 0.894 \\
\bottomrule
\end{tabular}
\end{table}
A quantitative comparison between the temperature anomaly fields obtained with the TDRW model and the COMSOL finite-element simulations (Figure~\ref{Fig3}(a)) is reported in Table~\ref{Tab2}. For weakly heterogeneous fractures ($\sigma_a/\langle a\rangle \le 0.21$), the agreement is excellent: the normalized root-mean-square error (NRMSE) remains below 0.03 and the coefficient of determination satisfies $R^2>0.99$. In this regime, the flow field is nearly uniform and heat transport is dominated by advection and longitudinal diffusion, which are both captured with high fidelity by the particle-based formulation. For the most heterogeneous case ($\sigma_a/\langle a\rangle = 0.60$), discrepancies increase moderately (NRMSE $\approx 0.1$, $R^2 \approx 0.89$), particularly in slow-flow regions where the TDRW tends to underestimate the thermal anomaly. These differences stem from the discrete, one dimensional coarse-grained representation of heat conduction inherent to the jump process: sub-grid temperature gradients that develop around nearly closed apertures in the FEM solution cannot be fully resolved by the particle scheme. Moreover, COMSOL treats heat conduction in the matrix with a three dimensional continuous PDE, whereas the TDRW evaluates fracture–matrix exchange via a semi-infinite convolution kernel applied at the walls. This leads to a slightly smoother temperature field in the vicinity of contact zones, although the overall structure of the thermal front and the dominant advective features remain very well reproduced.

Overall, the excellent agreement across the full range of heterogeneities confirms the ability of our TDRW scheme to reproduce the spatio-temporal evolution of the temperature field induced by fracture-scale variability. The TDRW scheme thus provides a 
reliable and computationally efficient surrogate for fully Eulerian simulations over a broad range of heterogeneity levels. Minor statistical fluctuations in the TDRW fields reflect the particle-based nature of the method and can be reduced further by increasing the number of walkers, albeit at higher computational cost.

\section{Quantitative Characterization of Heat Exchange Mechanisms}~\label{sec:analysis}
\subsection{Stochastic Framework}
To systematically investigate the interplay between longitudinal dispersion within fractures and conductive heat exchange with the surrounding matrix, we adopt a stochastic framework based on extensive Monte Carlo simulations. Specifically, we perform simulations across a range of realistic combinations of fracture geometry and flow parameters, as summarized in Table~\ref{Tab3}. We consider a total of 27 parameter combinations, defined by varying three key dimensionless parameters: the fracture closure ratio, the correlation ratio, and the Péclet number. The fracture closure ratio quantifies the variability in local fracture apertures; the correlation ratio characterizes the upper relative spatial scale for fracture heterogeneities (as compared to the fracture size) and determines both (i) the size of flow channeling patterns, and (ii) the degree of ergodicity of the velocity field; finally, the Péclet number captures the relative magnitude of advective  versus conductive heat transport rates.

The values for these three parameters were selected in this study to span the physically relevant regimes for advection–diffusion heat transport in rough fractures under Stokes flow. The thermal Péclet number is varied in the range $\Pen=10$–$100$, which corresponds to advection-dominated conditions in which fracture–matrix exchange is still significant enough to influence thermal breakthrough; they are relevant to geological fractures in the subsurface, among others. The roughness amplitude $\sigma_a/\langle a\rangle \in \{0.2,0.6,1.0\}$ captures the transition from mildly heterogeneous apertures to strongly constricted, partially
closed fractures. The correlation ratio $L/L_\mathrm{c}\in\{2,16,64\}$ spans
short- to long-range spatial organisation of aperture fluctuations,
reflecting the variability reported for natural and laboratory
fractures. Together, these ranges provide a representative set of
geometric and advective conditions while ensuring that all simulations
remain in the Stokes-flow regime and within the validity domain of the
semi-infinite matrix approximation.

Each Monte Carlo (MC) simulation set comprises $N_{\textrm{MC}}=100$ independent realizations to ensure robust statistical characterization of the inherent variability stemming from the heterogeneous aperture fields. These 100 realizations are generated for various sets of geometrical parameters defining a fracture geometry, i.e., the fracture closure ratio $\sigma_a/\langle a\rangle$, and the correlation ratios $L/L_\mathrm{c}$. 
For each fracture realization, fluid flow and heat transport are simulated. Transport simulations are run using a large ensemble of $N_{\textrm{p}}=10^6$ particles, which provides sufficient statistical resolution to accurately quantify the temporal evolution of the temperature field and to reliably estimate transport statistics such as the mean displacement, variance, and thermal exchange efficiency with the rock matrix.

In all such stochastic framework, the question of ergodicity, i.e., the statistical equivalence between
ensemble and spatial averages, such that a sufficiently large
realization of the aperture (and thus, flow) field may become representative of the underlying
stationary random process \citep{Dagan1989,Gelhar1993}, always arises. Here, for small fractures ($L/L_\mathrm{c}=2$), each fracture realization samples only a limited
number of independent correlated domains, leading to pronounced
realization-to-realization variability and reduced statistical
representativeness of each realization.  For larger $L/L_\mathrm{c}$ values, the number of independent aperture structures sampled within each domain is larger, and thus ensemble and spatial statistics of the
velocity field converge with increased $L/L_\mathrm{c}$, indicating a transition toward statistical ergodicity.  
This behaviour is consistent with the Monte Carlo analysis of flow and
transmissivity reported by \citet{Lenci2022b}, where increasing
$L/L_\mathrm{c}$ leads to a marked reduction in the dispersion of the PDFs of
velocity components, velocity magnitude, and effective transmissivity, showing
that sufficiently large fractures become statistically representative of the
medium.

In particular, \citet{Lenci2022b} showed that the velocity autocorrelation functions decay more rapidly in small systems, while the corresponding integral scales increase with domain size, reflecting reduced ensemble variability.  
\citet{Lenci2024} further quantified this transition by showing that the ratio between the correlation lengths of velocity and aperture fields approaches unity for $L/L_\mathrm{c}\!\gg\!1$, confirming the recovery of ensemble–spatial equivalence.  
Accordingly, increasing $L/L_\mathrm{c}$ enhances the statistical representativeness of individual realizations and ensures that macroscopic transport observables become invariant with domain size after normalization.

For each simulation set listed in Table~\ref{Tab3}, we evaluate three principal quantities of interest: (i) the mean longitudinal displacement, quantifying the overall advancement of the thermal front along the flow direction; (ii) the displacement variance, characterizing the spatial spreading of the thermal front and highlighting deviations from classical Fickian diffusion behavior; and (iii) the heat exchange efficiency at fracture–matrix interface. Both the Monte Carlo ensemble mean and the statistical variability, quantified through selected percentiles (e.g., the 5th and 95th percentiles), of these quantities are analyzed to provide robust insights into the transport processes and their uncertainty due to structural heterogeneity. These quantities, formally defined and analyzed in detail in subsequent sections, provide a comprehensive framework to investigate anomalous heat transfer regimes, their statistical variability, and their dependence on fracture and flow properties.
\begin{table}  
\centering
\caption{Monte Carlo combinations of roughness amplitude 
($\sigma_a/\langle a\rangle$), correlation ratio ($L/L_{\mathrm{c}}$), 
thermal Péclet number ($\Pen$), Reynolds number ($\Rey$), 
and imposed pressure gradient ($\langle \nabla P\rangle$) explored in this study.}
\label{Tab3}
\begin{tabular}{@{}lccccc@{}}
\toprule
\textbf{MC ID} & $\sigma_a/\langle a\rangle$ & $L/L_\mathrm{c}$ & $\Pen$ & $\Rey$ & $\langle \nabla P\rangle~(\textrm{Pa}/\mathrm{m})$\\
\midrule
MC 1  & 0.2 & 2  & 10  & 0.01 & 11  \\
MC 2  & 0.2 & 2  & 50  & 0.07 & 56  \\
MC 3  & 0.2 & 2  & 100 & 0.10 & 113 \\
MC 4  & 0.2 & 16 & 10  & 0.01 & 11  \\
MC 5  & 0.2 & 16 & 50  & 0.07 & 56  \\
MC 6  & 0.2 & 16 & 100 & 0.10 & 113 \\
MC 7  & 0.2 & 64 & 10  & 0.01 & 11  \\
MC 8  & 0.2 & 64 & 50  & 0.07 & 56  \\
MC 9  & 0.2 & 64 & 100 & 0.10 & 113 \\
MC 10 & 0.6 & 2  & 10  & 0.01 & 11  \\
MC 11 & 0.6 & 2  & 50  & 0.07 & 56  \\
MC 12 & 0.6 & 2  & 100 & 0.10 & 113 \\
MC 13 & 0.6 & 16 & 10  & 0.01 & 11  \\
MC 14 & 0.6 & 16 & 50  & 0.07 & 56  \\
MC 15 & 0.6 & 16 & 100 & 0.10 & 113 \\
MC 16 & 0.6 & 64 & 10  & 0.01 & 11  \\
MC 17 & 0.6 & 64 & 50  & 0.07 & 56  \\
MC 18 & 0.6 & 64 & 100 & 0.10 & 113 \\
MC 19 & 1.0 & 2  & 10  & 0.01 & 11  \\
MC 20 & 1.0 & 2  & 50  & 0.07 & 56  \\
MC 21 & 1.0 & 2  & 100 & 0.10 & 113 \\
MC 22 & 1.0 & 16 & 10  & 0.01 & 11  \\
MC 23 & 1.0 & 16 & 50  & 0.07 & 56  \\
MC 24 & 1.0 & 16 & 100 & 0.10 & 113 \\
MC 25 & 1.0 & 64 & 10  & 0.01 & 11  \\
MC 26 & 1.0 & 64 & 50  & 0.07 & 56  \\
MC 27 & 1.0 & 64 & 100 & 0.10 & 113 \\
\bottomrule
\end{tabular}
\end{table}

\subsection{Mean Displacement}
The longitudinal mean displacement is the average distance traveled by heat, advected by the flowing fluid, along the principal direction of the flow and is defined as 
\begin{equation}\label{eq:mean_disp}
\mathcal{M}(t)=\langle x_1(t) \rangle = \frac{1}{N_\textrm{p}} \sum_{i=1}^{N_\textrm{p}} x_1^{(i)}(t),
\end{equation}
where $x_1^{(i)}(t)$ is the position of the $i$-th heat particle at time $t$.

The mean displacement, commonly adopted in the literature, represents the first moment of the particle or heat displacement distribution and can be interpreted as the position of the center of mass of the thermal anomaly. Within the framework of advection–dispersion theory, the time derivative of the mean displacement yields the mean interstitial velocity, while higher-order moments such as the variance are used to derive effective longitudinal and transverse dispersion coefficients. Although the mean displacement alone does not capture the full extent of dispersive spreading, it remains a fundamental quantity due to its intrinsic connection with scalar conservation and the bulk advective flux.

The advancement of the thermal front is often associated with the point at which half of the initial temperature contrast is recovered, i.e., where $\Delta T_{\mathrm{f}}(\mathbf{x},t) = 0.5$. In systems with continuous injection and symmetric transport, this threshold may be interpreted as a proxy for the median of the particle displacement distribution. However, in strongly heterogeneous media, the displacement distribution becomes skewed, and the mean and median diverge \citep{Becker2000}. While the mean displacement can be more sensitive to extreme values, it remains the standard observable in transport modelling due to its direct connection to moment-based frameworks and its statistical robustness in TDRW approaches \citep{Dentz2004, Neuman2009}.

In a purely advective system with uniform velocity $u$, the mean displacement grows linearly with time:
\begin{equation}\label{eq:mean_disp_uniform}
\mathcal{M}(t) \approx ut,
\end{equation}
reflecting uniform thermal transport along the fracture. In realistic systems, however, additional mechanisms such as velocity-induced spreading, thermal diffusion, and conductive exchange with the surrounding matrix induce deviations from linearity and broaden the thermal front \citep{Zhou2022, Meng2023, Heinze2025}. As a result, the mean displacement captures the cumulative effects of all transport processes and highlights the transition from advection-dominated behavior to more complex regimes \citep{Dentz2004}.

In the limit of a uniform-aperture fracture, conductive exchange with the rock matrix introduces a broad distribution of trapping times due to diffusive excursions into the surrounding medium. At high Péclet numbers, advection dominates and trapping has limited effect, preserving the near-linear growth of the mean displacement. At low Péclet numbers, however, matrix conduction becomes significant: long trapping events accumulate and slow down the plume, with the mean displacement following a sublinear scaling, $\mathcal{M}(t) \propto t^{\alpha}$, where $\alpha$ approaches 0.5. This reflects the increasing influence of heavy-tailed trapping time distributions \citep{Haggerty1995,Zoia2010}.

In this work, we focus on the mean displacement as a classical metric to characterize thermal transport dynamics, in line with previous studies. While thermal fronts are often defined by specific temperature thresholds, the mean displacement remains a robust and widely used indicator of plume evolution. Anomalous transport is captured by a fractional scaling:

\begin{equation}\label{eq:mean_disp_anomalous}
\mathcal{M}(t) \propto t^{\alpha}, \quad \alpha \neq 1,
\end{equation}

where the exponent $\alpha$ reflects the interplay between advection, diffusion, and matrix conduction, leading to deviations from classical diffusive scaling.

In the context of heat transport in fractured media, deviations from linear displacement give rise to anomalous transport regimes, broadly classified as sub-ballistic or diffusion-dominated \citep{Berkowitz2006}. In sub-ballistic advective transport, the mean displacement grows faster than in classical diffusion but remains slower than pure advection, corresponding to a scaling
exponent $0.5 < \alpha < 1$. This regime is typically associated with preferential flow paths, where advective motion dominates over conductive losses for significant periods of time \citep{Moreno1993,Cortis2004,Bijeljic2006}. Conversely, subdiffusive transport arises when heat propagation is slower than expected from Fickian diffusion, with exponents $\alpha < 0.5$. This behavior is commonly attributed to long residence times in low-velocity zones or to strong conductive exchange with the surrounding matrix, but it may also emerge in purely advective systems due to extreme velocity contrasts caused by aperture variability within a single fracture \citep{Dentz2004, Fiori2015}. Classical diffusive behavior, corresponding to $\alpha = 0.5$, represents the limiting case where advection and dispersion are balanced by matrix conduction, resulting in plume spreading in accordance with standard diffusion laws. In this study, we use the term anomalous transport to refer to any departure from this diffusive scaling, with the exponent $\alpha$ providing a compact indicator of the underlying transport regime. This classification aids in interpreting the temporal evolution of the mean displacement across varying fracture heterogeneities, flow conditions, and spatial scales.

Notably, for a fixed fracture closure, the ensemble-averaged mean displacement $\mathcal{M}(t)$ remains statistically invariant with respect to the correlation ratio $L/L_\mathrm{c}$, reflecting scale-independent macroscopic transport properties. This invariance arises because ensemble averaging over multiple stochastic realizations effectively cancels finite-size variability, ensuring consistent transport behavior across different domain sizes.
The influence of $L/L_\mathrm{c}$ primarily manifests in the temporal window over which different transport regimes can be resolved. Small fractures, characterized by shorter streamlines and closer outlet boundaries, do not allow for a complete characterization of late-time behavior, as this would require particles to remain within the fracture domain for sufficiently long periods. Conversely, early-time discrepancies emerge from differences in mesh resolution. In smaller domains, finer meshes lead to shorter particle jumps and smaller temporal increments in the TDRW simulation, which enables a more refined characterization of short residence times and the early onset of transport regime transitions. In contrast, the coarser meshes used in larger domains result in larger spatial and temporal steps, which limit the resolution of early-time dynamics and reduce the number of particles contributing to the mean displacement at very short times.
In principle, a similar early-time characterization could be achieved in larger fractures by adopting the same fine mesh used for smaller domains. However, in our case, this would require a refinement by a factor of up to 30, leading to prohibitive computational costs for Monte Carlo simulations involving thousands of realizations. For this reason, while small differences at early times can be attributed to resolution effects, the scaling exponents governing both early and late regimes remain robust and statistically invariant across scales for a given closure.
\begin{figure}
\centerline{\includegraphics[width=1\textwidth]{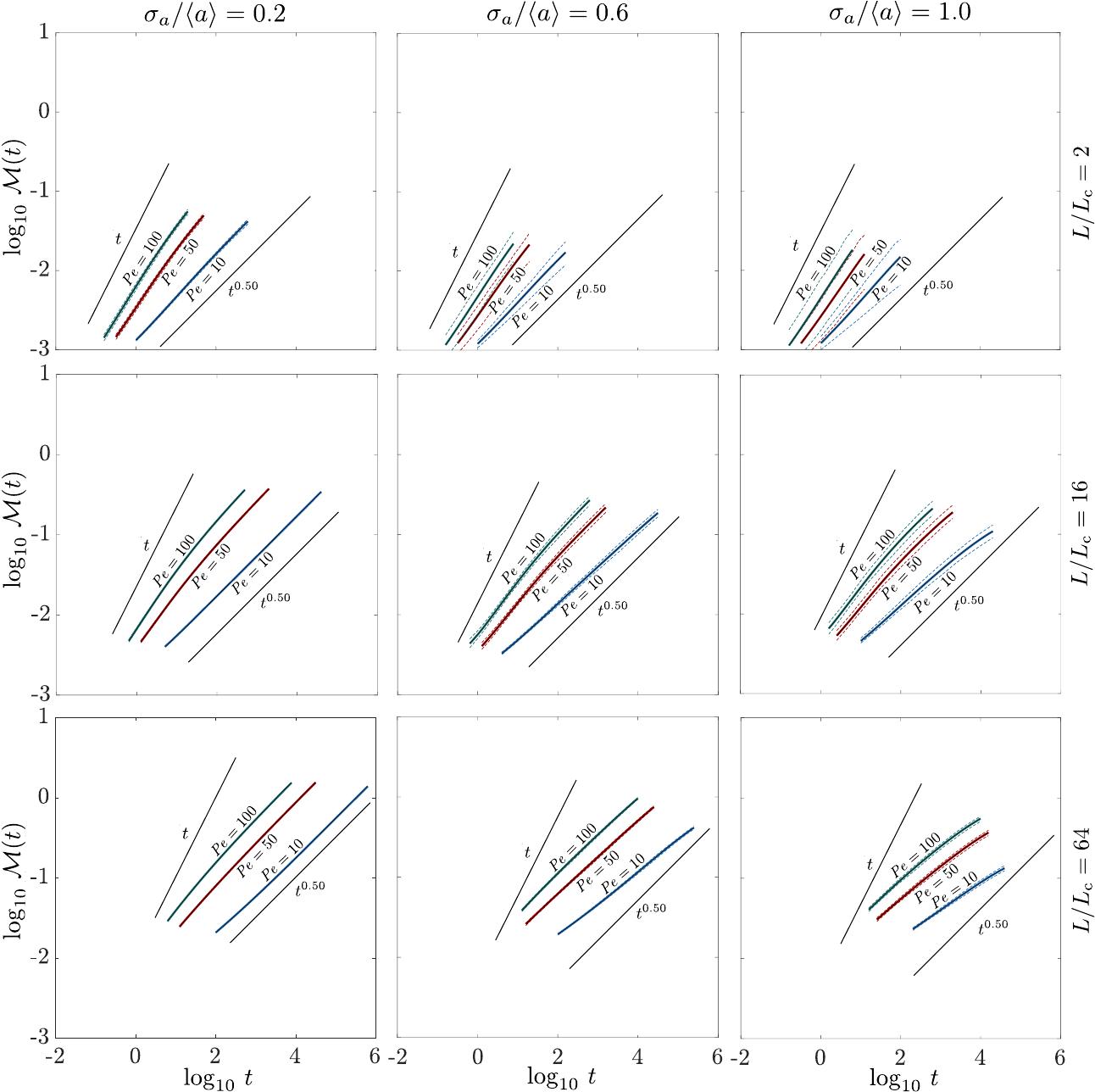}}
\caption{Temporal evolution of the ensemble averages (solid lines) and percentile bands (dashed lines) of the logarithm of the mean displacements, $\log_{10}{\mathcal{M}(t)}$, obtained from the Monte Carlo simulations described in Table~\ref{Tab3}. Each column, from top to bottom, corresponds to increasing fracture closures, $\sigma_a/\langle a\rangle = \{0.2,\,0.6,\,1.0\}$, reflecting growing aperture variability. Rows correspond to increasing ratios $L/L_\mathrm{c} = \{2,\,16,\,64\}$, indicating progressively larger fracture sizes relative to the correlation length, and hence increasing statistical ergodicity of the aperture field. Solid colored lines denote different Péclet numbers: $\Pen = 10$ (blue), $\Pen = 50$ (red), and $\Pen = 100$ (green). Black solid lines are included as visual references for diffusive ($\propto t^{0.5}$) and ballistic ($\propto t$) scaling. Mean displacement and time are expressed in meters and seconds, respectively. Simulation parameters are summarized in Table~\ref{Tab1}. Dashed lines indicate the 5th and 95th percentiles of the confidence interval.}
\label{Fig5}
\end{figure}

Figure~\ref{Fig5} shows the temporal evolution of the ensemble-averaged mean longitudinal displacement obtained from the Monte Carlo simulations summarized in Table~\ref{Tab3}. Solid lines represent the ensemble mean across all realizations, while dashed lines indicate the 5th and 95th percentiles, capturing the statistical variability associated with the stochastic fracture geometries.
Each column represents a different level of aperture variability, defined by the fracture closure values $\sigma_a/\langle a\rangle = 0.2$, $0.6$, and $1.0$, progressing from left (low heterogeneity) to right (high heterogeneity). Each row corresponds to increasing values of the correlation ratio $L/L_\mathrm{c} = 2$, $16$, and $64$, which reflect progressively larger fracture sizes, $L = 0.2\,\mathrm{m}$, $1.6\,\mathrm{m}$, and $6.4\,\mathrm{m}$, relative to the fixed correlation length $L_\mathrm{c} = 0.1\,\mathrm{m}.$

The figure highlights three distinct temporal regimes. At early times, particle motion is primarily governed by advection along preferential flow paths induced by aperture heterogeneity, as alternative transport mechanisms require sufficient time to become effective. This initial regime reflects the uniform injection of particles across the inlet, resulting in an initially homogeneous spatial distribution before the effects of flow heterogeneity manifest. As transport progresses, the particle distribution becomes increasingly structured due to velocity contrasts, with particles localizing within high-flow channels while others accumulate in quasi-stagnant zones. This reorganization marks the transition from an initially uniform distribution to a flux-weighted one. At later times, the combined influence of velocity-induced spreading and conductive exchange with the matrix broadens the particle plume and dominates the evolution of the thermal front.
For weak heterogeneity, the mean displacement exhibits non-ballistic advective scaling with exponents $0.5 < \alpha < 1$ for $\Pen = 50$ and $100$, whereas for $\Pen = 10$ the evolution approaches the diffusion-controlled limit ($\alpha \approx 0.5$).
This transition toward diffusive scaling at lower Péclet numbers arises because reduced advective velocities allow heat to remain longer within each voxel, enhancing fracture–matrix conductive exchange and accelerating the onset of diffusion-controlled behaviour.

Increasing the correlation ratio $L/L_{\mathrm{c}}$ does not alter the intrinsic flow tortuosity in this low-variability regime; instead, it increases the physical domain size and thus enlarges the temporal window available before particles reach the outlet.
Consequently, late-time diffusive behaviour becomes more clearly observable in the largest systems ($L/L_{\mathrm{c}} = 64$). Moreover, larger domains improve statistical ergodicity, reducing realization-to-realization variability (shown by the dotted-lines indicating the statistical confidence intervals) without modifying the ensemble-averaged mean displacement, which remains invariant across all values of $L/L_{\mathrm{c}}$.

For intermediate aperture variability ($\sigma_a/\langle a\rangle = 0.6$), the case with $L/L_{\mathrm{c}} = 2$ corresponds to a fracture whose length is only marginally larger than the correlation length.
In this regime, the small domain size increases the sensitivity to local aperture fluctuations, producing stronger realization-to-realization variability.
Nevertheless, the ensemble-averaged mean displacement remains statistically invariant across realizations, as expected from Monte Carlo averaging over stochastic heterogeneities. For larger $L/L_{\mathrm{c}}$ values, particles remain within the fracture for longer times before reaching the outlet, thereby extending the temporal window over which transport dynamics can be observed.
This enlarged window makes the early-time approach toward the diffusion-controlled limit more evident for $\Pen = 50$ and $100$.
For $\Pen = 10$, the mean displacement exhibits a pronounced late-time slowdown driven by strong matrix–conduction effects. 
The reduction in ensemble variability with increasing $L/L_{\mathrm{c}}$ reflects improved statistical ergodicity in larger domains, rather than any change in the underlying transport mechanisms.

At the highest aperture variability considered ($\sigma_a/\langle a\rangle = 1.0$), realization-to-realization variability is strongly amplified, particularly for small $L/L_{\mathrm{c}}$.
This increased dispersion reflects the well-known sensitivity of flow to severe aperture constrictions \citep{meheustJGR2001}. 
As $L/L_{\mathrm{c}}$ increases, the larger domain enhances statistical ergodicity, thereby reducing ensemble variability, although the ensemble-averaged mean displacement remains invariant with respect to $L/L_{\mathrm{c}}$, confirming that macroscopic transport is scale-independent.

At large closure, the combined action of strong aperture constrictions and extended matrix–conduction effects produces a pronounced advective slowdown, with mean-displacement exponents decreasing to $\alpha \approx 0.3$.
This marked reduction is driven by the dominance of long matrix-trapping events and extreme flow localization: heat is channeled through a small number of highly transmissive pathways, while extensive regions of the fracture remain quasi-stagnant.
These features substantially hinder the downstream progression of the thermal plume, yielding the most subdiffusive behaviour observed across all parameter sets.

\subsection{Displacement Variance}
The displacement variance is the second cumulant of the spatial distribution of heat particles along the flow direction,
\begin{equation} \label{Var_Discrete}
\mathcal{V}(t) = \frac{1}{N_\textrm{p}} \sum_{i=1}^{N_\textrm{p}} \left[x_1^{(i)}(t)\right]^2 - \left( \frac{1}{N_\textrm{p}} \sum_{i=1}^{N_\textrm{p}} x_1^{(i)}(t) \right)^2,
\end{equation}
and provides a quantitative measure of the spreading of the thermal anomaly over time. It captures the extent to which individual particle trajectories deviate from the mean position, and serves as a key indicator of thermal dispersion in heterogeneous systems. Analyzing the displacement variance allows one to infer the nature of the underlying transport regime, whether it follows normal diffusion, displays retardation effects, or exhibits enhanced spreading. This quantity is also directly linked to the effective longitudinal dispersion coefficient through
\begin{equation}
D^\mathrm{eff}_\mathrm{L}(t) = \frac{1}{2} \frac{d\mathcal{V}(t)}{dt},
\end{equation}
which characterizes the rate of plume broadening within the fracture \citep{Berkowitz2006, Chevalier1999, deDreuzyJGR2012}. 

In homogeneous systems subject only to advection at a constant velocity $u$, all particles move coherently, leading to negligible variance. When particle velocities $u_i$ vary but remain statistically stationary with finite variance, the displacement variance grows ballistically:
\begin{equation} \label{eq:var_disp}
\mathcal{V}(t) \propto t^2,
\end{equation}
which is typical of advection-dominated systems and reflects persistent velocity contrasts among particle paths.

For uniform-aperture fractures, early-time transport is characterized by advection and in-plane diffusion, producing ballistic behavior ($\mathcal{V}(t) \propto t^2$). As time progresses, in-plane diffusion gives $\mathcal{V}(t)\propto t$. At longer times, heat exchange with the matrix induces long residence times and a transition toward subdiffusive scaling ($\beta<1$), with heavy-tailed release governed by the semi-infinite diffusive memory of the matrix.

In heterogeneous systems, where the velocity field varies due to aperture fluctuations or geometric irregularities, the displacement variance reflects the combined influence of several transport processes \citep{Fox2015}. These include in-plane heat conduction, velocity-induced dispersion, and fracture–matrix exchange, all of which contribute to enhanced spreading and larger variance values \citep{Wang2023a}. At early times, the system typically exhibits a ballistic regime ($\mathcal{V}(t) \propto t^2$), associated with coherent advection before diffusive effects take over. This regime is generic in particle-based models and, while present even at low Péclet numbers, it is often unresolved due to its short duration. As thermal conduction and local velocity gradients become effective, the system transitions to near-Fourier scaling ($\mathcal{V}(t) \propto t$), eventually reaching subdiffusive behavior driven by matrix exchange and long residence times. 

At longer times, conductive exchange with the matrix acts as a retardation mechanism, while velocity-induced spreading continues to influence particle motion. The interplay between these processes often leads to anomalous transport characterized by non-linear scaling of the mean and variance, typically of the form:
\begin{equation} \label{eq:var_disp_anomalous}
\mathcal{V}(t) \propto t^{\beta}, \quad \beta \ne 1,
\end{equation}
where $\beta > 1$ indicates superdiffusion and $\beta < 1$ signals subdiffusion due to persistent retention in the matrix or structural trapping \citep{Dentz2012}. In most realistic fracture systems, the variance does not sustain ballistic growth over extended periods, reflecting the increasingly complex interplay between heterogeneous
advection in the fracture plane, in-plane thermal conduction within the fluid, and fracture–matrix exchange driven by matrix conduction.
\begin{figure}
\centerline{\includegraphics[width=1\textwidth]{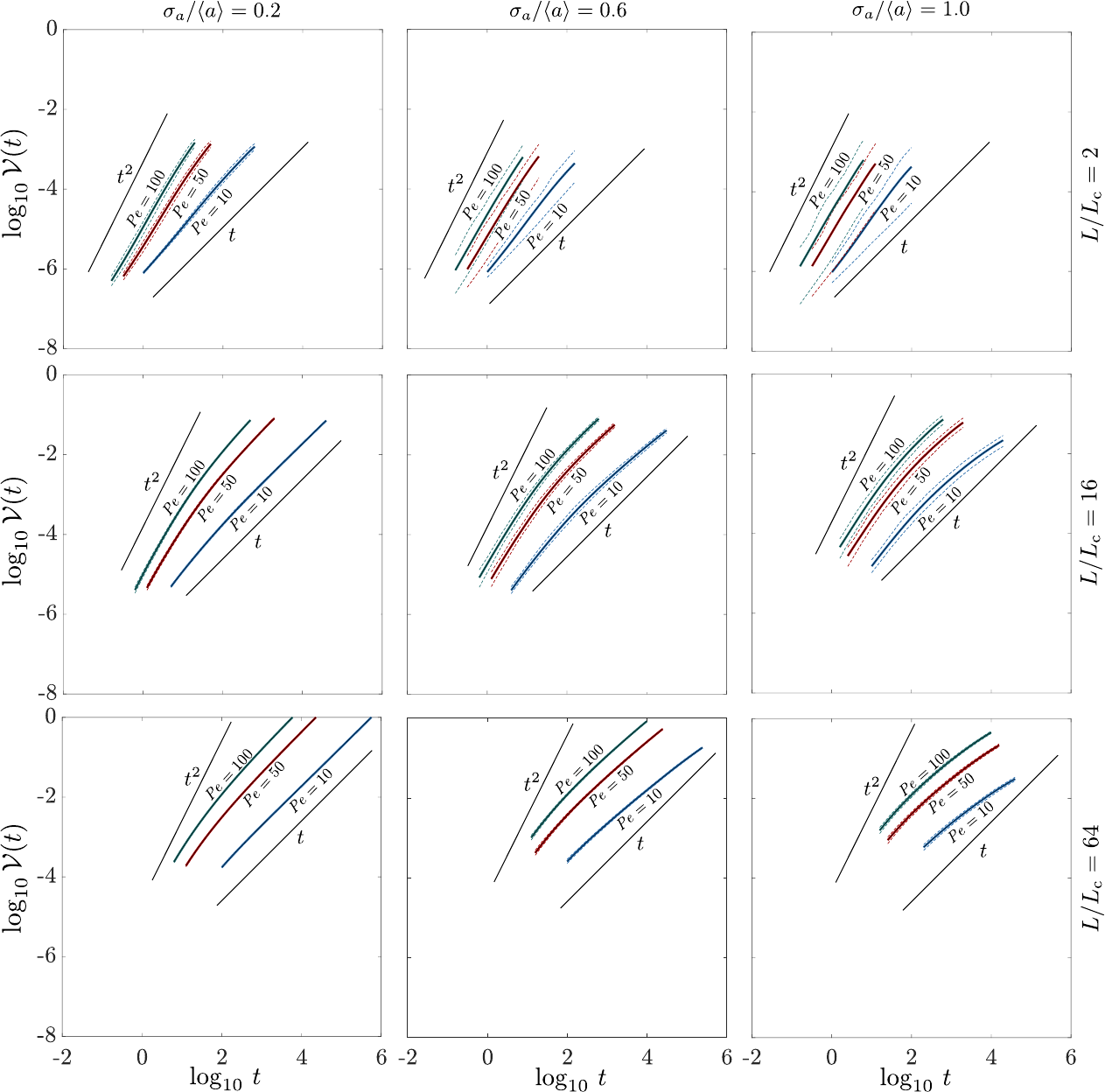}}
\caption{
Temporal evolution of the ensemble-averaged displacement variance, $\log_{10}\mathcal{V}(t)$, obtained from the Monte Carlo simulations summarized in Table~\ref{Tab3}. 
Solid lines denote ensemble means, while dashed lines represent the 5th–95th percentile range across realizations. 
Columns (left to right) correspond to increasing fracture closures, $\sigma_a/\langle a\rangle = \{0.2,\,0.6,\,1.0\}$; rows (top to bottom) show increasing correlation ratios, $L/L_\mathrm{c} = \{2,\,16,\,64\}$. 
Colored curves indicate different Péclet numbers: $\Pen = 10$ (blue), $\Pen = 50$ (red), and $\Pen = 100$ (green). 
Black reference lines indicate diffusive scaling ($\mathcal{V}\propto t$, slope~1 in log–log) and ballistic scaling ($\mathcal{V}\propto t^{2}$, slope~2). Variance and time are expressed in $\mathrm{m}^2$ and $\mathrm{s}$, respectively.}
\label{Fig6}
\end{figure}

Figure~\ref{Fig6} presents the temporal evolution of the displacement variance $\mathcal{V}(t)$ for the Monte Carlo simulations described in Table~\ref{Tab3}. As in the case of the mean displacement (Figure~\ref{Fig5}), each panel shows the ensemble average (solid line) along with the 5th and 95th percentiles (dashed lines), highlighting both the central trend and the statistical variability across realizations.
Each column corresponds to increasing aperture heterogeneity, as measured by the fracture closure ratio $\sigma_a/\langle a\rangle = 0.2$, $0.6$, and $1.0$, while each row reflects an increase in the correlation ratio $L/L_\mathrm{c} = 2$, $16$, and $64$. These parameters control, on the one hand, the intensity of local velocity contrasts, and, on the other hand, the spatial ergodicity of the aperture field as well as the spatial extent of flow channeling patterns.
At early times, the displacement variance exhibits clear superdiffusive behavior across most configurations, with scaling exponents significantly greater than unity. This reflects the dominance of advective transport along preferential channels, which is especially pronounced at high Péclet numbers. For low closure and short correlation length ($\sigma_a/\langle a\rangle = 0.2$, $L/L_\mathrm{c} = 2$), the displacement variance exhibits nearly ballistic growth, with $\mathcal{V}(t) \propto t^2$, reflecting persistent velocity contrasts between flow paths in the absence of significant diffusive smoothing. This behavior is particularly evident at high Péclet numbers, where advective channeling sustains a clear superdiffusive or ballistic scaling over extended times. At low Péclet numbers, the same initial trend is present but persists over much shorter time scales, often falling below the resolution of the simulation. This is because the transition from ballistic to diffusive transport occurs more rapidly as diffusion dominates earlier. A more complete comparison across correlation ratios could be achieved by normalizing time by the characteristic advective time required to travel a correlation length. Increasing the correlation ratio amplifies these effects, leading to stronger early-time spreading due to enhanced channeling. However, in the largest domains ($L/L_\mathrm{c}=64$), the early-time dynamics begin to moderate, particularly at low Péclet, where the influence of heat conduction into the matrix becomes more visible.
At late times, a transition toward diffusive or subdiffusive behavior is observed, depending on the interplay between fracture heterogeneity and matrix conduction. For low heterogeneity, the variance tends to approach a diffusive regime, especially for large correlation ratios. This reflects the increasing influence of matrix exchange, which acts as a retardation mechanism and dampens the early superdiffusive trends. As the fracture closure increases, late-time behavior becomes increasingly subdiffusive, with lower scaling exponents indicating the dominance of long residence times in stagnant zones and more effective heat transfer into the rock matrix.
A key observation is that the statistical dispersion across realizations increases with aperture heterogeneity (higher closure), due to more pronounced local velocity contrasts and flow localization. However, this variability tends to decrease with increasing correlation ratio, as longer fracture domains promote more ergodic behavior and reduce the relative influence of local anomalies, as was the case for the mean plume displacement, as discussed above. In highly heterogeneous yet large-scale systems, ensemble behaviour converges more reliably toward average trends. Nevertheless, strong variability is seen between individual realizations for the same statistical geometrical parameters

Overall, the behavior of the displacement variance reveals a consistent picture of early-time spreading dominated by preferential flow and velocity heterogeneity, followed by late-time regimes shaped by matrix conduction. The reduction in variability across realizations observed in larger domains reflects enhanced ergodicity, which leads to more consistent statistical behavior. This makes displacement variance a robust diagnostic for identifying the dominant heat transfer mechanisms in fractured media.

While the ensemble-averaged variance remains statistically invariant across different values of $L/L_\mathrm{c}$ for a given closure, minor deviations are observed in both the early-time and late-time regimes. At early times, the differences arise primarily from the finer spatial discretization in smaller fractures, which enables shorter jumps and thus higher temporal resolution within the TDRW framework. This enhanced resolution allows for a more accurate representation of the initial spreading driven by velocity variability. At late times, however, shorter fractures limit the residence time of particles within the domain, impeding the full development of the subdiffusive regime governed by conductive exchange with the matrix. As a result, larger fractures provide a more complete characterization of the variance evolution at late times. Nevertheless, the scaling exponents governing the growth of the variance remain consistent across different $L/L_\mathrm{c}$ values when ensemble-averaged, confirming that aperture variability (closure) is the dominant control factor, while the fracture size primarily affects the temporal extent over which different regimes can be observed. These results confirm that while fracture size ($L/L_\mathrm{c}$) controls both the temporal window over which different transport regimes can be observed and the statistical variability across individual realizations, since larger domains average over more heterogeneities, it is the aperture variability (closure) that fundamentally governs the nature and scaling of thermal dispersion.

\subsection{Breakthrough curves and late-time behaviour}
The late--time behaviour of breakthrough curves is analysed by focusing on the
complementary cumulative distribution function of arrival times,
$CCDF(t)=1-BTC(t)$, which directly quantifies the fraction of heat that remains
within the fracture--matrix system at time $t$. This representation is particularly well suited to identifying slow release mechanisms, since algebraic decay of the $CCDF$ provides a direct signature of long-lived trapping and nonlocal memory effects, whereas such behaviour is progressively obscured by the saturation of $BTC$ as $t\to\infty$.

Monte Carlo simulations show that the long--time decay of $CCDF$ is not
universal, but depends on the relative importance of advective transport and
conductive exchange with the surrounding matrix. In advection--dominated regimes (high Péclet numbers and weak aperture heterogeneity), $CCDF$ decays rapidly and remains close to exponential over the time window accessible to the simulations, reflecting a relatively narrow distribution of residence times. In contrast, when velocity contrasts and matrix conduction become significant, the $CCDF$ exhibits a slow algebraic decay,
\begin{equation}
CCDF(t) \propto t^{-\gamma},
\end{equation}
with an exponent $\gamma$ that reflects the balance between advective transport
and diffusive retention. Equivalently, the corresponding breakthrough curve approaches its asymptotic value as $BTC(t) \simeq 1 - t^{-\gamma}$, highlighting the persistent contribution of late--arriving particles to the thermal release.

At high Péclet numbers, heat transfer is dominated by advection and the decay
remains steep, whereas at lower $\Pen$ values conductive retention becomes
increasingly important and $\gamma$ decreases below~1.
The diffusive limit, corresponding to transient conduction into a semi--infinite
matrix, yields $\gamma=0.5$, consistent with the long--time behaviour of the
survival probability associated with a Lévy--Smirnov kernel.
This differs from the classical $t^{-3/2}$ decay of outlet breakthrough curves
following an impulse injection, which corresponds to the probability density of
exit times rather than to the survival function considered here.

These theoretical limits provide useful benchmarks to interpret the simulation
results presented below, where the progressive broadening and flattening of
$CCDF$ with increasing heterogeneity and fracture size reflect the continuous
transition between advective, diffusive, and anomalous transport regimes.
\begin{figure}
\centerline{\includegraphics[width=1\textwidth]{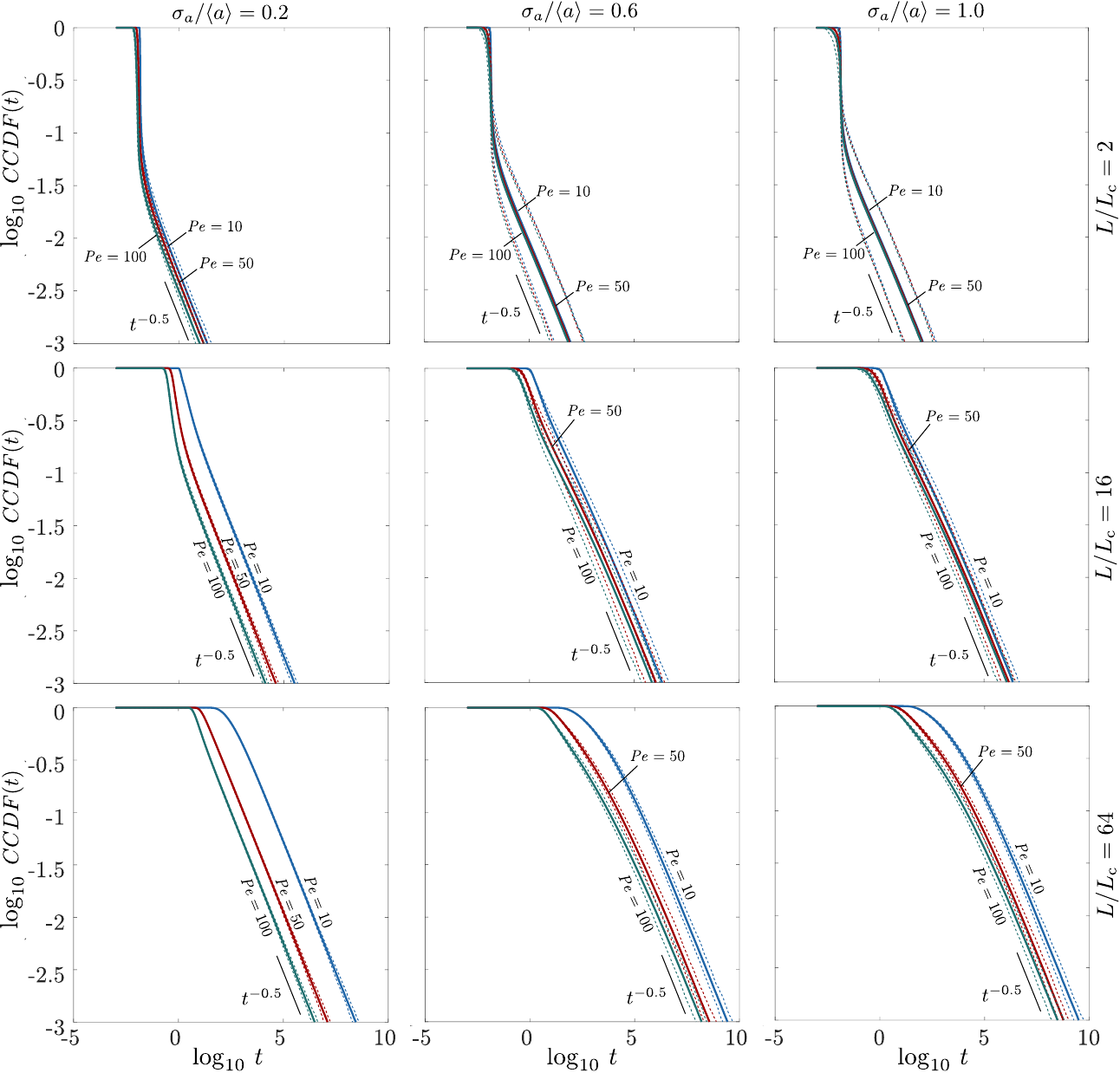}}
\caption{Temporal evolution of the ensemble averages (solid lines) and percentile bands (dashed lines) of the complementary cumulative distribution of arrival times, $\log_{10}CCDF(t)$, obtained from the Monte Carlo simulations summarized in Table~\ref{Tab3}. 
Each column (top to bottom) corresponds to increasing fracture closures, $\sigma_a/\langle a\rangle = \{0.2,\,0.6,\,1.0\}$, representing progressively stronger aperture variabilities. 
Rows correspond to increasing correlation ratios $L/L_\mathrm{c} = \{2,\,16,\,64\}$, which determine how extensively each realization samples the underlying aperture heterogeneity and thus the range of residence times accessible within the fracture. Solid coloured lines indicate different Péclet numbers: $\Pen = 10$ (blue),
$\Pen = 50$ (red), and $\Pen = 100$ (green). The decay of $CCDF$ quantifies the fraction of heat remaining within the fracture–matrix system over time. Early-time rapid decay reflects advective breakthrough through high-conductance channels, whereas intermediate shoulders and late-time heavy tails arise from velocity heterogeneity and conductive exchange with the matrix. Increasing fracture closure produces slower decay and more pronounced tailing, while larger $L/L_\mathrm{c}$ values reduce variability across realizations by providing a more representative sampling of the aperture field and by extending the temporal window over which distinct transport regimes can be observed. Time is expressed in seconds.}
\label{Fig9}
\end{figure}

Figure~\ref{Fig9} shows the complementary cumulative distributions of arrival
times, $CCDF(t)=1-BTC(t)$, obtained from the Monte Carlo simulations for
increasing levels of aperture heterogeneity and correlation ratio
$L/L_\mathrm{c}$. Although the physically relevant observable for heat recovery is the
breakthrough curve $BTC$ under continuous injection, its complement
$CCDF$ is analysed here because it provides a more sensitive diagnostic of
late--time dynamics. In particular, algebraic tails associated with conductive trapping and
nonlocal memory effects are directly visible in $CCDF$, whereas they are
progressively obscured by the saturation of $BTC$ as $t\to\infty$.
Each curve represents the fraction of heat that has not yet reached the outlet
at time~$t$, and therefore remains stored within the fracture--matrix system.
Its temporal decay directly reflects the competition between advective transport
along preferential flow paths and conductive retention within low--velocity
regions and the surrounding rock matrix.

For low aperture variability ($\sigma_a/\langle a\rangle = 0.2$, left column),
the $CCDF$s decay sharply after a short initial transient, indicating that most
of the heat exits the fracture rapidly through nearly uniform, high-velocity
flow paths. The corresponding distribution of travel times is relatively
narrow, with only a modest late-time fraction associated with conductive
exchange with the matrix. This regime is dominated by advection and mild longitudinal dispersion, as conductive exchange with the rock plays a secondary role.
The slight dependence on the Péclet number observed at early times reflects the relative contribution of conduction: at lower $\Pen$, heat spends a longer duration in contact with the fracture walls, which slightly increases the fraction of delayed particles and smoothes the early-time front, whereas at higher $\Pen$ advection dominates and the $CCDF$ decays more abruptly. In all cases, the late-time slope approaches $t^{-0.5}$, consistently with a diffusive regime governed by transient conduction in a semi-infinite matrix.

As the closure increases to $\sigma_a/\langle a\rangle = 0.6$ and $1.0$ (center and right columns), the breakthrough curves become markedly broader and exhibit more pronounced heavy tails. 
Increasing fracture closure amplifies aperture contrasts, generating high-conductance channels separated by low-aperture zones with extremely small velocities. 
This enhances both velocity-induced spreading and the residence time of heat within stagnant regions, promoting repeated conductive exchange with the matrix. 
The $CCDF$s therefore transition smoothly from the sharp exponential-like decay of the advective regime to a slower, algebraic decline extending over several decades in time.

The onset of the heavy tail marks the time scale at which matrix conduction
begins to control the release of heat stored in low‐velocity and quasi-stagnant regions. In this transition regime, the apparent slope of $CCDF$ may become
slightly flatter than the asymptotic $t^{-1/2}$ behaviour, reflecting the broad
distribution of residence times generated by the coupling between
fracture-scale velocity heterogeneity and intermittent diffusive exchange with
the matrix. Such pre-asymptotic flattening is consistent with observations from
direct numerical simulations in rough fractures
\citep[e.g.][]{deSimone2021,deSimone2023}, where heterogeneous advection and
local trapping produce intermediate-time tails that are heavier than the
diffusive limit.

At the highest closure, $\sigma_a/\langle a\rangle = 1.0$, these effects are
most pronounced: the $CCDF$s develop extended linear segments in the log–log
representation, revealing well-defined power-law behaviour before ultimately
converging to the classical $t^{-1/2}$ decay dictated by semi-infinite matrix
diffusion. The persistence of these long residence times indicates strong
non-Fickian retention, with a finite fraction of heat remaining trapped in the
system over very long durations.

The ratio $L/L_\mathrm{c}$ controls the spatial representativeness of the aperture field and, consequently, the statistical stability of the $CCDF$s. Furthermore, the total duration of the breakthrough is limited, as a smaller fracture allows fewer long residence times to develop before heat exits the system.
As $L/L_\mathrm{c}$ increases, the fracture samples many correlation patches,
making each realization more representative of the ensemble statistics and
yielding smoother, more stable $CCDF$s. The longer domain also allows access to
longer residence times, thereby revealing a broader range of transport regimes.

Overall, the analysis of breakthrough behaviour reveals that fracture closure governs the degree of non-Fickianity by setting the strength of velocity contrasts and the prevalence of conductive trapping, whereas the correlation ratio $L/L_\mathrm{c}$ controls the statistical stability and temporal extent of the observed regimes.

\subsection{Fracture–Matrix Heat Exchange Efficiency}
The fracture--matrix coupling derived in Section~\ref{sec:transport} yields a temporally nonlocal interfacial flux with kernel $(t-\tau)^{-1/2}$ (Eqs.~\eqref{flux_q_dim}–\eqref{B_def}).  
In the TDRW formulation, the aperture--averaged fracture temperature anomaly is reconstructed from particle arrivals,
\begin{equation}
\Delta T_{\mathrm{f}}(\mathbf{x},t)=F(\mathbf{x},t),
\end{equation}
where $F(\mathbf{x},t)$ is the cumulative fraction of heat--carrying particles that have reached $\mathbf{x}$ by time $t$.  
Differentiation over time gives
\begin{equation}
\label{eq:dDelta_f_to_dt}
\frac{\partial \Delta T_{\mathrm{f}}}{\partial t}(\mathbf{x},t)
=
\frac{\partial F}{\partial t}(\mathbf{x},t)
=
f(\mathbf{x},t),
\end{equation}
where $f(\mathbf{x},t)$ is the associated arrival--time probability density function (PDF). The thermal anomaly within the matrix can be described using Duhamel’s principle. This approach constructs the solution to the time-dependent problem as a temporal convolution of the $\operatorname{erfc}$ kernel with the fracture temperature history, effectively superposing instantaneous solutions weighted by the rate of change of the interface temperature \citep{Duhamel1837}:
\begin{equation}\label{eq:matrix_temperature}
\Delta T_{\textrm{m}}(x_3^\prime,t)=\int_0^t \frac{\partial \Delta T_{\textrm{f}} (\mathbf{x},\tau)}{\partial\tau}\operatorname{erfc}\left[\frac{x_3^\prime}{2\sqrt{\alpha_\textrm{m}(t-\tau)}}\right]\, d\tau.
\end{equation}
Substituting Eq.~\eqref{eq:dDelta_f_to_dt} into Eq.~\eqref{flux_q_dim} yields the particle--based representation of the interfacial heat flux
\begin{equation}\label{eq:q_kernel_pdf_link}
q_{\mathrm{m}}(\mathbf{x},t)
=
\frac{2\,\phi_{\mathrm{m}} k_{\mathrm{r}}(T_1-T_0)}{\sqrt{\pi \alpha_{\mathrm{m}}}}
\int_0^{t}
\frac{f(\mathbf{x},\tau)}{\sqrt{t-\tau}}\,d\tau,
\end{equation}
highlighting that fracture–matrix heat exchange results from the convolution of a particle arrival‐time PDF with the fractional conductive kernel. Expressing $q_{\mathrm{m}}$ in terms of the arrival–time PDF makes the
equivalence with the deterministic Volterra formulation explicit:
the TDRW summation becomes the Monte–Carlo analogue of the
fractional convolution kernel. In this representation, the PDF
encodes both advective transport and Lévy–distributed matrix
residence times, while the $(t-\tau)^{-1/2}$ kernel provides the
semi–infinite conduction response. This confirms that the stochastic
model recovers the same memory operator as the classical
Duhamel–Volterra solution.

The convolution is accumulated on the fly during particle propagation, without storing the full arrival history.
In discrete form, if $\tau_{k}^{(i,j)}$ denotes the $k$th particle--arrival time at control volume $(i,j)$, the interfacial flux becomes
\begin{equation}\label{eq:q_discrete}
q_{\mathrm{m}}(\mathbf{x}_j,t)
\simeq 
\frac{2\,\phi_{\mathrm{m}} k_{\mathrm{r}}(T_1-T_0)}
     {\sqrt{\pi \alpha_{\mathrm{m}}}}
\sum_{\tau_k(\mathbf{x}_j)<t}
\frac{1}{\sqrt{t-\tau_k(\mathbf{x}_j)}},
\end{equation}
which is the stochastic analogue of the Volterra convolution in Eq.~\eqref{B_def}.  

The total heat transferred to the matrix follows as
\begin{equation}\label{eq:thermal_power_clean}
P_{\mathrm{m}}(t)
\;\approx\;
\frac{2\,\phi_{\mathrm{m}}k_{\mathrm{r}}(T_1-T_0)\,\Delta x^2}{\sqrt{\pi\,\alpha_{\mathrm{m}}}}
\sum_{\mathbf{x}_j\in\Omega_{\mathrm{f}}}
\sum_{\tau_k(\mathbf{x}_j)<t}
\frac{1}{\sqrt{t-\tau_k(\mathbf{x}_j)}}\,.
\end{equation}
where the outer sum spans all fracture control volumes and the inner sum accumulates contributions from arrivals within each volume.  
The factor $2$ accounts for heat exchange at both fracture walls.  
This incremental evaluation preserves the exact Lévy--Smirnov memory structure implied by the semi--infinite matrix without post--processing.

If the interface temperature were instead imposed instantaneously (i.e., without Duhamel’s principle), the fracture would relax exponentially, $P_{\mathrm{m}}(t)\propto e^{-t/\tau}$, as in memoryless single--timescale models.  
In contrast, Eqs.~\eqref{eq:q_kernel_pdf_link}--\eqref{eq:thermal_power_clean} yield the characteristic algebraic decay $P_{\mathrm{m}}(t)\propto t^{-1/2}$, reflecting the persistent thermal memory of a semi--infinite matrix and the nonlocal nature of fracture--matrix heat exchange.

\begin{figure}
\centerline{\includegraphics[width=1\textwidth]{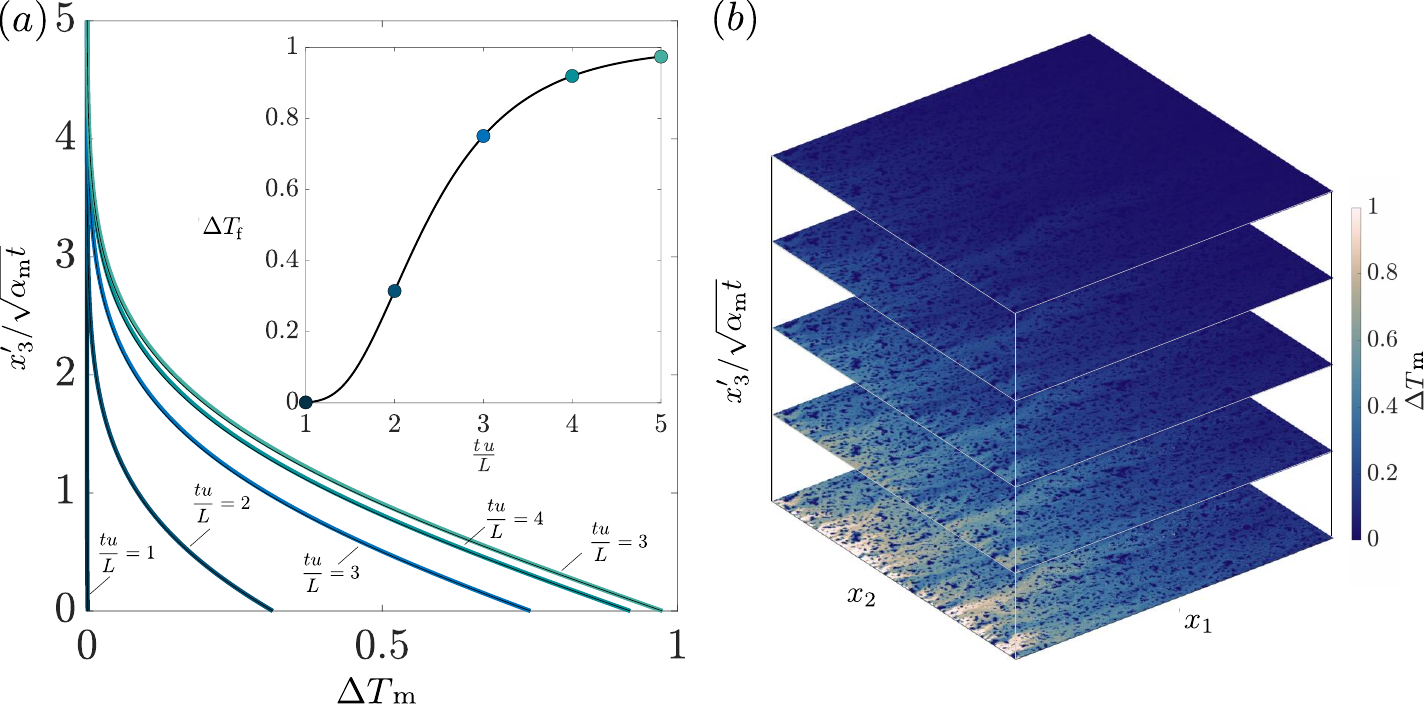}}
\caption{Panel (a) shows the variation of the matrix thermal anomaly $\Delta T_{\textrm{m}}$ as a function of the dimensionless matrix depth $x_3^\prime/\sqrt{\alpha_\textrm{m}t}$. The inset displays the time evolution of the thermal anomaly $\Delta T_{\mathrm{f}}$ at the interface $x_3^\prime = 0$, i.e., within the fractures  as given by Eq.~\eqref{eq:CDF_arrival_times}. Panel (b) illustrates the matrix thermal anomaly at different depths across planes parallel to the fracture–matrix interface ($x_3^\prime = 0$).}
\label{Fig4}
\end{figure}
Figure~\ref{Fig4} illustrates the evolution of the thermal anomaly in the matrix domain due to conductive exchange with the fractures.  
Panel~(a) presents the dimensionless temperature profiles $\Delta T_{\textrm{m}}$ as a function of the rescaled matrix depth $x_3^\prime/\sqrt{\alpha_\textrm{m}t}$. This self-similar representation highlights the diffusive nature of heat propagation into the matrix, showing that temperature gradients evolve consistently across different times when appropriately normalized. The inset shows the evolution of the fracture temperature $\Delta T_{\mathrm{f}}$ at the interface ($x_3^\prime = 0$), obtained from Eq.~\eqref{eq:CDF_arrival_times} is transferred to the matrix.  
Panel~(b) further examines the matrix thermal response by showing $\Delta T_{\textrm{m}}$ across several planes parallel to the fracture–matrix interface. These cuts at increasing depths provide a spatial perspective on the thermal penetration, confirming the expected monotonic decay away from the interface and emphasizing the limited spatial extent of the temperature front at early times.

From the total thermal power exchanged at the fracture–matrix interface, we define the thermal efficiency $\varepsilon(t)$ as the ratio between the instantaneous thermal power transferred to the matrix, $P_{\mathrm{m}}(t)$, and the thermal power injected into the fracture at the inlet boundary, $P_{\textrm{inj}}$, thus providing a dimensionless measure of the effectiveness of heat extraction:
\begin{equation}\label{thermal_efficiency}
\varepsilon(t)=\frac{P_{\mathrm{m}}(t)}{P_{\textrm{inj}}}.
\end{equation}
where the inlet thermal power is expressed as $P_{\textrm{inj}}=\rho_{\textrm{w}}\,c_{p,\textrm{w}}\,Q\,(T_1-T_0)$, where $Q$ is the volumetric flow rate through the fracture, thereby linking thermal efficiency directly to the fluid flow and heat input conditions.

\begin{figure}
\centerline{\includegraphics[width=1\textwidth]{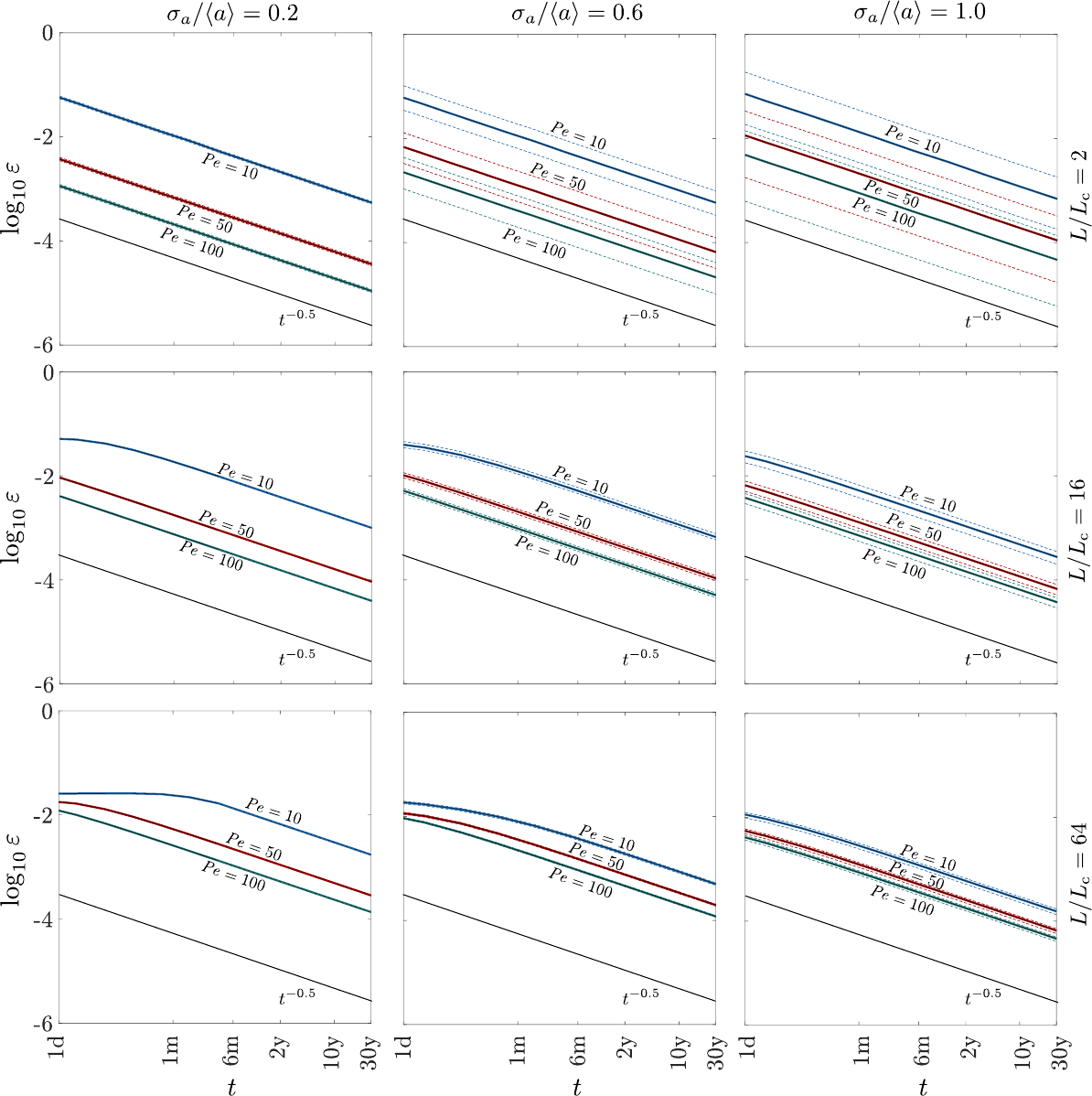}}
\caption{
Temporal evolution of the ensemble averages (solid lines) and percentiles (dashed lines) of the logarithm of the heat exchange efficiency, $\log_{10}(\varepsilon)$, obtained from the Monte Carlo simulations described in Table~\ref{Tab3}. 
Each column (top to bottom) corresponds to increasing fracture closures, $\sigma_a/\langle a \rangle = \{0.2,\,0.6,\,1.0\}$, i.e. increasing aperture variability. 
Rows correspond to increasing correlation ratios, $L/L_\mathrm{c} = \{2,\,16,\,64\}$. 
Solid colored lines indicate different Péclet numbers: $\Pen = 10$ (blue), $\Pen = 50$ (red), and $\Pen = 100$ (green). 
Black solid lines serve as visual references for the diffusive behavior in a semi-infinite matrix ($\propto t^{-0.5}$). 
Time is expressed in seconds.
}
\label{Fig7}
\end{figure}

Figure~\ref{Fig7} shows the time evolution of the thermal power $P_\mathrm{m}(t)$ exchanged across the fracture–matrix interface for the 27 Monte Carlo combinations. Each panel displays the ensemble average (solid line) and the 5th and 95th percentiles (dashed lines), quantifying variability due to aperture field heterogeneity. At sufficiently long times (depending on the fracture closure, characteristic length ratio and Péclet number), the observed decay of thermal efficiency follows a power-law behavior, $\propto t^{-1/2}$ at sufficiently long times, which is a hallmark of diffusive heat transfer into a semi-infinite medium. This behavior reflects a fundamental physical property: in such geometries, the matrix retains thermal memory, and heat penetrates progressively rather than instantaneously. In our formulation \eqref{eq:q_kernel_pdf_link}, this feature is not imposed a priori, but rather emerges naturally from the convolution with a memory kernel derived from the solution of the heat equation in the matrix. Specifically, the Lévy--Smirnov kernel embodies the nonlocal-in-time response of the matrix to a thermal perturbation applied at the fracture--matrix interface. This kernel encodes the long-tailed memory inherent to the system and enables an accurate representation of the temporal evolution of heat exchange. In contrast, local-in-time (memoryless) models cannot reproduce this algebraic decay, because they neglect the persistence of thermal gradients in the matrix and therefore predict an exponential relaxation. The observed power-law behaviour thus confirms the necessity of a temporally non-local kernel to capture the essential physics of fracture–matrix heat exchange.

For a low fracture closure ($\sigma_a/\langle a\rangle = 0.2$) and small correlation ratio ($L/L_\mathrm{c} = 2$), the thermal power scales as $t^{-1/2}$ for all Péclet numbers, reflecting the classical solution of diffusive exchange with a semi-infinite matrix. The heat flux is relatively uniform across the fracture, as evidenced by the near-complete overlap between percentile bands and the ensemble mean. This indicates that all realizations exhibit similar flow patterns and thermal behavior due to the limited heterogeneity. Thermal power is significantly higher at $\Pen=10$, exceeding that of $\Pen=100$ by over two orders of magnitude. This occurs because lower advective velocities give more time for heat exchange with the matrix, enhancing thermal efficiency.

Increasing the fracture closure to $\sigma_a/\langle a\rangle = 0.6$ (for $L/L_\mathrm{c}=2$) results in a marked increase in mean thermal power, especially for $\Pen=100$, which shows an order-of-magnitude gain. The $t^{-1/2}$ scaling persists, but the spread between realizations grows, as indicated by the widening percentile bands. This variability likely stems from increased aperture heterogeneity and stronger flow channeling, which enhance residence time contrasts and promote more localized heat exchange with the matrix. In contrast, at lower Péclet numbers ($\Pen=10$), although the advective flow field still reflects the same heterogeneity, conductive heat transport within the matrix becomes more dominant relative to advection. This reduces the sensitivity of the system to localized velocity fluctuations by smoothing temperature gradients at the fracture–matrix interface. As a result, the spatial variability in heat uptake is attenuated, and the mean thermal power remains nearly unaffected by changes in fracture closure. In other words, matrix conduction acts as a buffer that limits the impact of channelization on overall heat exchange efficiency.

At the highest heterogeneity level ($\sigma_a/\langle a\rangle = 1.0$), the mean power increases further while maintaining the $t^{-1/2}$ trend. However, the inter-realization variability becomes much more significant, up to an order of magnitude, highlighting the strong sensitivity of heat exchange to local flow organization. This dispersion reflects the combined effects of elevated aperture variability and a limited domain size relative to the correlation length, reducing statistical averaging and increasing stochastic sensitivity.

Increasing the correlation ratio to $L/L_\mathrm{c} = 16$ leads to a more ergodic system, as larger fracture domains sample a broader spectrum of aperture fluctuations. This enhances statistical averaging along individual realizations and reduces the dispersion around the ensemble mean. For $\sigma_a/\langle a\rangle = 0.2$, the average behavior consistently follows the expected $t^{-1/2}$ scaling at long times across all Péclet numbers. At $\Pen = 10$, a distinct initial plateau emerges at early times ($t < 1$ month), indicating that heat exchange initiates rapidly and is maintained at nearly constant efficiency over a finite duration. This regime reflects the thermal inertia of the matrix, which temporarily limits the rate at which additional energy can be conducted into the rock, despite the relatively weak influence of advective transport in the fracture. The tight overlap between percentiles and the ensemble mean further confirms the spatial uniformity of heat uptake under conditions of low aperture variability.

As closure increases to $\sigma_a/\langle a\rangle = 0.6$, the thermal power curves show greater variability and an earlier departure from the $t^{-1/2}$ scaling, particularly at low $\Pen$. For $\sigma_a/\langle a\rangle = 1.0$, the mean thermal efficiency decreases across all Péclet numbers, but the effect is strongest for $\Pen = 10$ and minimal for $\Pen = 100$. This behavior reflects a faster onset of matrix saturation under stronger heterogeneity, which limits heat uptake efficiency. At low $\Pen$, where conduction dominates, the matrix saturates more easily, leading to shorter and weaker thermal exchange phases. In contrast, at high $\Pen$, advective spreading delays saturation and sustains heat transfer efficiency.

At the highest correlation ratio ($L/L_\mathrm{c} = 64$), the system includes significantly longer fractures, which leads to broader spatial domains for heat exchange. As a result, the statistical dispersion around the ensemble mean is reduced, yielding narrower confidence intervals across realizations. For $\sigma_a/\langle a\rangle = 0.2$, the $\Pen = 10$ case displays a pronounced plateau lasting up to 6 months, whereas it is much less evident for $\Pen = 50$ and $\Pen = 100$. The extent of this plateau increases with larger fracture length, lower closure (and thus, reduced aperture heterogeneity), and lower Péclet number. These conditions promote more uniform and sustained heat transfer: the flow is less channelized, the fracture surfaces are more homogeneously active, and particles experience longer residence times, allowing for prolonged exchange with the matrix. As in the $L/L_\mathrm{c} = 16$ case, this plateau reflects the thermal inertia of the matrix and the time required for the temperature gradient to decline significantly. The matrix initially absorbs heat efficiently due to the strong contrast with the fracture temperature, and this phase persists until the gradient diminishes enough for the exchange rate to decline and the system enters the classical diffusive regime. It is important to note that the flow field is stationary, but thermal transport remains transient, since the fracture temperature evolves over time as the thermal front propagates. Thus, the plateau does not indicate a delay in the onset of exchange, but rather a transient phase of sustained and nearly constant heat transfer, governed by the slow thermal response of the matrix.

At long times, all cases converge to the characteristic $t^{-1/2}$ decay, consistent with diffusive uptake in a semi-infinite medium. 
While initial thermal power is similar across Péclet numbers, the $\Pen = 10$ case ultimately achieves higher efficiency, with power levels nearly an order of magnitude greater than those observed at higher flow rates.

Overall, these results illustrate that the system is most thermally efficient under slow and spatially uniform flow conditions (low $\Pen$ and low $\sigma_a/\langle a\rangle$), while strong aperture heterogeneity and large fracture domains lead to increased variability across realizations. Nonetheless, the $t^{-1/2}$ scaling characteristic of matrix diffusion remains robust at long times across most configurations, confirming that diffusive exchange governs the asymptotic thermal behavior.

\subsection{Computational workflow}
\label{sec:flowchart}
\begin{figure}
  \centering
  \includegraphics[width=\textwidth]{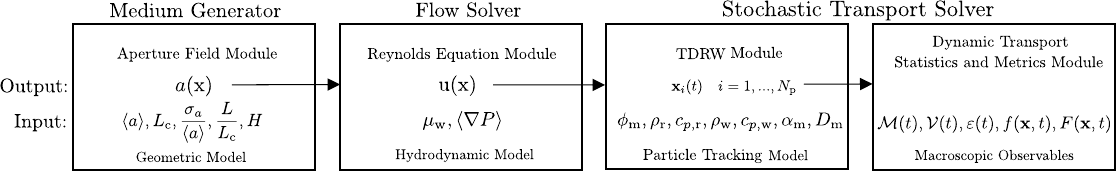}
  \caption{Computational workflow linking aperture generation, lubrication-flow solution, and stochastic particle transport. 
  Modules: (1) Aperture Field Generator, (2) Reynolds Solver, (3) TDRW Transport, (4) Dynamic Transport Statistics.}
  \label{fig:fig8}
\end{figure}

The overall workflow is shown in figure~\ref{fig:fig8}, highlighting the sequential coupling between geometric characterisation, hydrodynamic solution, and stochastic transport. Four modules are involved, each associated with a specific physical model and numerical procedure.
The first module, the Aperture Field Generator, produces a truncated self-affine Gaussian aperture field (Section~\ref{sec:medium}), which prescribes the spatial variability of the local aperture for the subsequent flow computation. 
The synthesis procedure follows the spectral algorithm described in \cite{Lenci2022a}.
The second module, the Reynolds Solver (Section~\ref{sec:flow}), evaluates steady-state lubrication flow through the variable-aperture fracture by solving the Reynolds equation for pressure and velocity fields. 
The finite–volume discretisation and numerical implementation are consistent with the methodology reported in \cite{Lenci2022a}.
The remaining two modules constitute the stochastic transport framework.  
The TDRW Transport Module (Section~\ref{sec:transport}) advances particles according to the time-domain random-walk formalism, combining advective displacements with the Lévy–Smirnov matrix-trapping times derived in Appendix~\ref{appA}. 
The Dynamic Transport Statistics Module (Section~\ref{sec:analysis}) evaluates macroscopic observables on-the-fly (e.g.\ breakthrough curves, residence-time distributions, matrix heat uptake, thermal efficiency), thereby eliminating the need for post-processing and enabling memory-efficient accumulation of statistics.  
Implementation details, computational scaling, and performance considerations are discussed in Appendix~\ref{appC}, including numerical considerations, memory-management strategy, and discussion of computational cost and algorithmic scaling.
This workflow provides a direct numerical realisation of the coupled fracture-matrix heat transport process described above, enabling efficient Monte Carlo estimation of fracture-scale thermal observables without further approximations.

\section{Conclusions}~\label{sec:conclusion}
This study presents a physically grounded stochastic framework for modelling heat transport in fractured geological media, where large-scale thermal dynamics are governed by the interplay between advection by the heterogeneous flow within rough-walled fractures, heat conduction within the fluid, conductive exchange with the surrounding low-permeability rock matrix, and heat conduction within that matrix. The proposed approach couples a Time-Domain Random Walk (TDRW) formulation with a semi-analytical model for fracture–matrix heat transfer, capturing both early-time transport and long-time anomalous retention within a unified and computationally efficient particle-tracking scheme.

A key element of the framework is the explicit representation of matrix trapping times through the conditioned Lévy–Smirnov distribution, derived from first-passage theory for transient heat conduction in a semi-infinite medium. This heavy-tailed law, with its characteristic $t^{-3/2}$ decay and infinite mean, follows directly from the one-dimensional heat equation with a fixed boundary temperature and does not contain any adjustable parameters. Unlike empirical Multi-rate Mass Transfer (MRMT) approaches or phenomenological memory-kernel models, the trapping-time statistics arise directly from the solution of the heat equation in a semi-infinite matrix and depend only on measurable quantities: the fracture aperture, the heat-capacity ratio, and the matrix thermal diffusivity. As a result, the fracture–matrix exchange law contains no tunable parameters and is fully predictive within the considered physical setting.

While the Lévy–Smirnov model effectively captures essential features of early-time trapping and long residence times resulting from matrix diffusion, its infinite mean limits its applicability in systems with finite spatial extent and intersecting fractures. It remains a suitable approximation in advection-dominated regimes, particularly when the Péclet number is greater than one and transport is primarily governed by fracture-scale dynamics. Incorporating a physical cutoff, for example by considering a finite diffusion time across the matrix, would lead to finite moments and improve realism in bounded domains. Such considerations have not been considered here and are left for future work.

Monte Carlo simulations across 27 parameter combinations highlight the influence of aperture heterogeneity ($\sigma_a/\langle a\rangle$), correlation ratio ($L/L_\mathrm{c}$), and Péclet number ($\Pen$) on coupled flow and heat transfer in fracture-matrix systems. The simulations demonstrate that non-Fickian thermal behaviour arises from the coupled action of heterogeneous advection and matrix conduction. Spatial organisation of the flow field determines the distribution of mobile residence times and therefore the duration of particle–wall contact.
Because the trapping-time distribution is conditioned on this advective residence, the Lévy–Smirnov law naturally encodes the interplay between velocity intermittency and conductive uptake.
Matrix conduction provides the intrinsic non-local memory, while flow heterogeneity modulates its expression along channelised pathways set by the aperture geometry.
Their combined effect produces clear transitions from superdiffusive or ballistic transport at early times, primarily driven by advective channeling within high-aperture pathways, to subdiffusive behavior at later times, where thermal conduction into the matrix becomes the dominant mechanism. These behaviors, observed 
 across the explored Péclet numbers, reveal a direct mechanistic link between fracture scale heterogeneity, advective organisation, and the emergent thermal signatures.
The variability observed across realizations increases with local heterogeneity and is larger for larger fractures (at fixed correlation length), emphasizing the inherent trade-off between disorder and ergodicity at the system scale.

While the ensemble mean remains finite and exhibits a linear scaling in high-Péclet conditions, the variance reveals a markedly different behavior: its persistent growth over time reflects the impact of long retention events in the matrix, especially under strong aperture heterogeneity. These trends confirm that classical Fickian transport models are inadequate to describe large-time behavior, even in simple single-fracture settings. Instead, the interplay between flow variability and matrix exchange induces anomalous transport signatures, with implications for both predictive modelling and experimental interpretation.

The analysis of breakthrough curves further corroborates the physical consistency of the model. The complementary cumulative distribution of arrival times, $CCDF$, encapsulates the fraction of thermal energy that remains stored within the fracture--matrix system as a function of time. Monte Carlo simulations reveal a progressive transition from sharp, exponential-like decay in nearly homogeneous fractures to broad, power-law tails in highly heterogeneous ones.  This evolution reflects the growing influence of conductive retention and long trapping times in the matrix, which impart a pronounced memory effect to the transport dynamics.  At late times, the $CCDF$s converge to an asymptotic scaling $t^{-1/2}$, consistent with transient conduction in a semi-infinite matrix and with the analytical behavior predicted by the memory-kernel formulation. These results confirm that the heavy-tailed release of heat arises naturally from the coupling between advective variability and diffusive storage, without invoking any empirical parameterization.

To compute the heat flux at the fracture–matrix interface, we have introduced a nonlocal convolution formulation based on Duhamel’s principle, which accounts for the full temporal memory of conductive transport in the matrix.
The local temperature history was convolved with a universal kernel proportional to $(t-\tau)^{-1/2}$, allowing the flux to be evaluated on the fly during TDRW simulations using only particle arrival times. This approach avoids matrix discretization, preserves the exact temporal weighting associated with semi–infinite diffusion, and recovers the asymptotic $t^{-1/2}$ decay in thermal power predicted by classical theory.
At early times, deviations from the asymptotic regime may appear as short
plateau--like trends in the interfacial heat flux. These arise from the strong
initial temperature contrast between the hot fluid and the cold matrix, which
sustains a nearly constant flux until the near–interface rock warms. During this
phase, the exchange is limited not only by the imposed temperature gradient, but
also by the matrix’s thermal inertia: its volumetric heat capacity controls how
much energy can be absorbed, while its thermal diffusivity governs how rapidly
heat redistributes within the solid. As the matrix warms and the interfacial
gradient relaxes, the flux naturally transitions to the $t^{-1/2}$ regime
characteristic of semi--infinite conduction.
Within the kernel formulation, this behaviour emerges directly from the
convolution with the $(t-\tau)^{-1/2}$ memory operator, which can be readily
generalized to account for finite--size effects, reactive or stratified media,
and anomalous diffusion processes.

In summary, the proposed framework provides a transparent, scalable, and physically interpretable approach for modelling heat transport in fractured media. By grounding both mobile and immobile phase dynamics in first-principles physics and avoiding ad hoc parameterization, it enables robust characterization of thermal exchange, and supports uncertainty quantification, upscaling, and performance assessment in subsurface energy systems. Future extensions may incorporate finite matrix size effects, complex fluid rheology, temperature-dependent viscosity, or full fracture networks. 

The model has allowed us to understand the multiscale competition between geometric heterogeneity, advecting flow localisation, and matrix diffusion of heat. The resulting dynamics reproduce the key non-Fickian signatures observed in fractured rock, including broad residence-time distributions, anomalous scaling, and persistent thermal contrasts, while clearly delineating the conditions under which semi-infinite matrix diffusion provides an accurate physical description. The stochastic analysis reveals a sequence of transport regimes across the explored heterogeneity range. At early times, heat propagation is primarily governed by advection through high-aperture, high-transmissivity channels induced by self-affine wall roughness, whereas matrix conduction progressively dominates the late-time dynamics. Fracture closure controls the balance between advective channeling and matrix trapping, thereby setting the transition time from superdiffusive to subdiffusive thermal transport. The correlation ratio primarily controls the spatial persistence of preferential pathways and enlarges the spatial and temporal window over which transport dynamics develop, while also modulating ensemble dispersion through realization-to-realization variability. Together, these geometric parameters determine both the efficiency of heat exchange and the scaling behaviour and statistical variability of thermal transport in rough-walled fractures.

\section*{List of symbols}

\begin{table}
  \centering
  \caption{Physical and geometric symbols used in the manuscript.}
  \label{tab:symbols1}
  \renewcommand{\arraystretch}{1.2}
  \footnotesize
  \begin{tabular}{@{}llp{0.58\textwidth}@{}}
  \toprule
  \textbf{Symbol} & \textbf{Unit} & \textbf{Description} \\
  \midrule
  
  $a(\mathbf{x})$ & m & Local fracture aperture at position $\mathbf{x}$ \\
  $a_\mathrm{m}$ & m & Mechanical aperture \\
  $\langle a\rangle$ & m & Mean aperture \\
  $\sigma_a$ & m & Standard deviation of aperture field \\
  $L$ & m & Fracture length \\
  $L_\mathrm{c}$ & m & Aperture correlation length \\
  $H$ & -- & Hurst exponent of fracture-wall roughness \\[1mm]

  $\mathbf{u}=(u_1,u_2)$ & m\,s$^{-1}$ & Depth‐averaged in‐plane velocity \\
  $\mathbf{j}$ & m$^2$\,s$^{-1}$ & Volumetric flux per unit width \\
  $\langle\nabla P\rangle$ & Pa\,m$^{-1}$ & Macroscopic pressure gradient \\
  $U_\mathrm{c}$ & m\,s$^{-1}$ & Characteristic velocity \\[1mm]

  $\rho_\mathrm{w}$ & kg\,m$^{-3}$ & Fluid density \\
  $\mu_\mathrm{w}$ & Pa\,s & Fluid viscosity \\
  $k_\mathrm{w}$ & W\,m$^{-1}$\,K$^{-1}$ & Thermal conductivity of the fluid \\
  $c_{p,\mathrm{w}}$ & J\,kg$^{-1}$\,K$^{-1}$ & Specific heat of the fluid \\[1mm]

  $\rho_\mathrm{r}$ & kg\,m$^{-3}$ & Density of the rock \\
  $k_\mathrm{r}$ & W\,m$^{-1}$\,K$^{-1}$ & Thermal conductivity of the rock \\
  $c_{p,\mathrm{r}}$ & J\,kg$^{-1}$\,K$^{-1}$ & Specific heat of the rock \\
  $\alpha_\mathrm{m}$ & m$^2$\,s$^{-1}$ & Thermal diffusivity of the rock \\[1mm]

  $\phi_\mathrm{r}$ & -- & Matrix porosity of the rock\\
  $\phi_\mathrm{m}$ & -- & Volumetric heat‐capacity ratio between matrix and fluid \\[1mm]

  $T_\mathrm{f},T_\mathrm{m}$ & K & Fluid and matrix temperatures \\
  $\Delta T_\mathrm{f}, \Delta T_\mathrm{m}$ & -- & Dimensionless fluid and matrix temperature anomalies \\
  $T_0,T_1$ & K & Initial and injection temperatures \\[1mm]

  $q_\mathrm{m}$ & W\,m$^{-2}$ & Conductive heat flux at the fracture–matrix interface \\
  $P_\mathrm{m}$ & W & Conductive power transferred to matrix \\
  $P_\mathrm{inj}$ & W & Injected thermal power \\
  $\varepsilon(t)$ & -- & Thermal exchange efficiency \\[1mm]

  $\Rey$ & -- & Reynolds number \\
  $\Pen$ & -- & Thermal Péclet number \\

  \bottomrule
  \end{tabular}
\end{table}

\begin{table}
  \centering
  \caption{Transport, stochastic, and numerical symbols used in the manuscript.}
  \label{tab:symbols2}
  \renewcommand{\arraystretch}{1.2}
  \footnotesize
  \begin{tabular}{@{}llp{0.60\textwidth}@{}}
  \toprule
  \textbf{Symbol} & \textbf{Unit} & \textbf{Description} \\
  \midrule

  $\mathcal{M}(t)$ & m & Mean longitudinal displacement \\
  $\mathcal{V}(t)$ & m$^2$ & Variance of longitudinal displacement \\
  $BTC(t)$ & -- & Breakthrough curve \\[1mm]
   $CCDF(t)$ & -- &  Complementary cumulative distribution of arrival times \\[1mm]
  $D^\mathrm{eff}_\mathrm{L}$ & m$^2$\,s$^{-1}$ & Effective longitudinal thermal dispersion coefficient \\[1mm]

  $\alpha, \beta, \gamma$ & -- & Scaling exponents ($\mathcal{M}\!\sim\! t^\alpha$, 
    $\mathcal{V}\!\sim\! t^\beta$, $CCDF\!\sim\! t^{-\gamma}$) \\[1mm]

  $\theta_\mathrm{f}$ & s & Mobile residence time in the fracture \\
  $\theta_\mathrm{m}$ & s & Matrix trapping (immobile) residence time \\[1mm]

  $\psi_\mathrm{f}(t)$ & s$^{-1}$ & PDF of $\theta_\mathrm{f}$ (exponential) \\
  $\psi_\mathrm{m}(t)$ & s$^{-1}$ & PDF of $\theta_\mathrm{m}$ (Lévy--Smirnov) \\[1mm]

  $A_j$ & s$^{1/2}$ & Coupling coefficient $2\phi_\mathrm{m}\sqrt{\alpha_\mathrm{m}}\,\tau_j/a_j$ \\
  $\tau_j$ & s & Mean advective residence time in cell $j$ \\[1mm]

  $P_{j\to i}$ & -- & Transition probability from cell $j$ to neighbour $i$ \\
  $b_{ij}$ & s$^{-1}$ & Finite-volume flow-exchange coefficient \\
  $V_i, S_{ij}$ & m$^3$, m$^2$ & Volume of cell $i$; area of face $(i,j)$ \\
  $\sigma(i)$ & -- & Set of neighbors of cell $i$ \\[1mm]

  $\Delta x$ & m & Grid spacing \\
  $N_\mathrm{MC}$ & -- & Number of Monte Carlo realizations \\
  $N_\mathrm{p}$ & -- & Number of particles per realization \\
  $N_\mathrm{mesh}$ & -- & Number of grid cells in the domain \\[1mm]

  $\nabla,\,\nabla^2$ & -- & Gradient and Laplacian operators \\
  $\operatorname{erfc}(\cdot)$ & -- & Complementary error function \\

  \bottomrule
  \end{tabular}
\end{table}

\backsection[Acknowledgements]{Part of the computing for this project was performed on the Sherlock cluster. The authors gratefully acknowledge Stanford University and the Stanford Research Computing Center for providing computational resources and support that contributed to these research results. We also acknowledge ISCRA for awarding access to the Leonardo supercomputer, owned by the EuroHPC Joint Undertaking and hosted by CINECA (Italy). The research activities were carried out at the Department of Energy Science and Engineering of Stanford University, within the Data-Driven modelling and Simulations Group. The Department of Civil, Chemical, Environmental, and Materials Engineering (DICAM), University of Bologna, serves as the beneficiary institution and coordinates the GEONEAT project. Géosciences Rennes, CNRS Université de Rennes 1, through the TERA (Fluids, Transport and Reactivity) research team, actively collaborated in the scientific development of the project and will also host a secondment in accordance with the fellowship’s implementation plan.}

\backsection[Funding]{"This project has received funding from the European Union's Horizon Europe research and innovation program under the Marie Sklodowska-Curie grant agreement No. 101111216, Project GEONEAT — Complex Fluids in Fractured Geological Media for Enhanced Heat Transfer". Part of the computational resources were provided by CINECA through an ISCRA Class C allocation on the Leonardo supercomputer.  Views and opinions expressed are, however, those of the authors only, and do not necessarily reflect those of the European Union or the European Research Council Executive Agency. Neither the European Union nor the granting authority can be held responsible for them."}

\backsection[Declaration of interests]{The authors report no conflict of interest.}

\backsection[Data availability statement]{All data supporting the findings of this study are available within the figures presented in the paper, which are derived from numerical solutions of the equations discussed. There are no restrictions on data availability.}

\backsection[Author ORCIDs]{Lenci, 0000-0002-0285-6991; Méheust, 0000-0003-1284-3251; Klepikova, 0000-0003-4290-2400; Di Federico, 0000-0001-9554-0373; Tartakovsky, 0000-0001-9019-8935.}

\backsection[Author contributions]{
A. Lenci conceived the study, developed the methodology, implemented the software, performed the simulations, analyzed the results, curated the data, created the visualizations, wrote the original draft, and contributed to funding acquisition.  
Y. Méheust contributed to the conceptual framework, methodology, formal analysis, supervision, writing, and funding acquisition.  
M. Klepikova contributed to software implementation, validation, data curation, and writing.  
V. Di Federico contributed to the conceptual development, supervision, writing, and conducted the project administration and funding acquisition.  
D.M. Tartakovsky contributed to conceptual development, methodology, supervision, writing, project administration, and funding acquisition.  
All authors reviewed and edited the manuscript.
}

\appendix

\section{Derivation of the Matrix Trapping-Time Distribution}\label{appA}

The matrix trapping-time distribution originates from the classical one-dimensional model of heat exchange between a flowing fluid and an adjacent porous matrix, first introduced by \citet{Lauwerier1955}. In this system, heat is transported by advection along the fracture and by diffusion, along the direction transverse to the mean fracture plane, into the rock. The fracture is assumed planar, of uniform aperture $a$. The governing equations read as
\begin{align}
\frac{\partial T_{\mathrm{f}}}{\partial t}
+ u\,\frac{\partial T_{\mathrm{f}}}{\partial x}
&=
-\,\frac{2\,\phi_{\mathrm{m}}}{a}\,
\alpha_{\mathrm{m}}\,
\frac{\partial T_{\mathrm{m}}}{\partial z}\bigg|_{z=\pm a/2},
&& x\ge 0,
\label{eq:lauw1}\\[4pt]
\frac{\partial T_{\mathrm{m}}}{\partial t}
&=
\alpha_{\mathrm{m}}\,
\frac{\partial^2 T_{\mathrm{m}}}{\partial z^2},
&& x\ge 0,\; |z|\ge \tfrac{a}{2}.
\label{eq:lauw2}
\end{align}
where \(u\) is the mean fluid velocity, $x$ the spatial coordinate along the fracture flow direction, and $z$ the spatial coordinate along the direction transverse to the fracture plane. 
Equations~\eqref{eq:lauw1} and~\eqref{eq:lauw2} are subject to the following initial and boundary conditions:
\begin{equation}
T_{\mathrm{f}}(x,0)=T_0, \qquad T_{\mathrm{m}}(x,z,0)=T_0, \qquad
T_{\mathrm{f}}(0,t)=T_1, \qquad
T_{\mathrm{f}}(x\to\infty,t)=T_0, \qquad
T_{\mathrm{m}}(x,\pm\infty,t)=T_0 .
\end{equation}
Local thermal equilibrium at the fracture--matrix interface is assumed, so that
continuity of temperature at the fracture walls gives
\begin{equation}
\label{eq:temp_continuity}
T_{\mathrm{m}}(x,\pm\tfrac{a}{2},t)=T_{\mathrm{f}}(x,t).
\end{equation}
Solving~\eqref{eq:lauw1}–\eqref{eq:lauw2} subject to~\eqref{eq:temp_continuity}
yields the classical Lauwerier solution
\begin{equation}\label{eq:lauwerier_profile}
\Delta T_{\mathrm{f}}(x,t)
\equiv 
\frac{T_{\mathrm{f}}(x,t)-T_0}{T_1-T_0}
=
\operatorname{erfc}\!\left(
\frac{\phi_{\mathrm{m}}\sqrt{\alpha_{\mathrm{m}}}\,x}
     {a\,u\,\sqrt{t - x/u}}
\right),
\qquad t > x/u .
\end{equation}
Introducing the local advective contact time
\begin{equation}
\label{eq:theta_f_def}
\theta_{\mathrm{f}} = \frac{x}{u},
\end{equation}
equation~\eqref{eq:lauwerier_profile} can be rewritten as
\begin{equation}
\label{eq:zeta_def}
\Delta T_{\mathrm{f}}(\theta_{\mathrm{f}},t)
=
\operatorname{erfc}\!\left(
\frac{\phi_{\mathrm{m}}\sqrt{\alpha_{\mathrm{m}}}\,\theta_{\mathrm{f}}}
     {a\,\sqrt{t-\theta_{\mathrm{f}}}}
\right).
\end{equation}
In the classical solution of Lauwerier, heat is transported advectively along the fracture and reaches the position $x$ after the advective time $\theta_{\mathrm{f}}$. The temperature at $x$ depends on the elapsed time since
the arrival of the advective front, namely on the residual time $\theta_{\mathrm{m}} = t - \theta_{\mathrm{f}}$, which represents the time available for conductive heat diffusion into the surrounding rock matrix.

To obtain a stochastic analogue of the Eulerian solution, we exploit the fact that the normalized fracture temperature anomaly $\Delta T_{\mathrm{f}}\in(0,1]$ is a bounded and monotonic function of time. It can therefore be interpreted as a cumulative distribution function in an equivalent stochastic formulation. Let $\eta \sim \mathcal U(0,1]$ denote a uniformly distributed random variable. Introducing the auxiliary variable $Z=\operatorname{erfc}^{-1}(\eta)$ and inverting equation~\eqref{eq:zeta_def} with respect to $\theta_{\mathrm{m}}$ yields
\begin{equation}\label{eq:tdiff_def}
\theta_{\mathrm{m}}
=
\left(
\frac{\phi_\mathrm{m}\sqrt{\alpha_\mathrm{m}}\,\theta_{\mathrm{f}}}
     {a\,\operatorname{erfc}^{-1}(\eta)}
\right)^{2}.
\end{equation}
Equation~\eqref{eq:tdiff_def} defines a sampling rule for the random matrix trapping time $\theta_{\mathrm{m}}$. The associated probability density function is obtained by standard change-of-variables arguments, yielding 
\begin{equation}\label{eq:phi_j}
\psi_{\mathrm{m}}(t)
= \sqrt{\frac{C}{\pi\,t^3}}\,
  \exp\!\left(-\frac{C}{t}\right),
  \qquad
  C = \left(\frac{\phi_\mathrm{m}\sqrt{\alpha_\mathrm{m}}\,
  \theta_{\mathrm{f}}}{a}\right)^{2}.
\end{equation}
This is the Lévy–Smirnov distribution:
\begin{equation}
\psi_{\mathrm{m}}\!\left(t \,\big|\, \theta_{\mathrm{f}}\right)
= \frac{\phi_{\mathrm{m}}\sqrt{\alpha_{\mathrm{m}}}\,\theta_{\mathrm{f}}}
       {a\sqrt{\pi}}\,
  t^{-3/2}\!
  \exp\!\left[
    -\,\frac{\phi_{\mathrm{m}}^{2}\,\alpha_{\mathrm{m}}\theta_{\mathrm{f}}^{2}}
           {a^{2}t}
  \right],
  \qquad t>0.
\end{equation}
The heavy-tailed factor \(\propto t^{-3/2}\) reflects the broad distribution of residence times associated with diffusion into an effectively semi-infinite matrix. 

\section{Residence-time distribution and equivalent memory kernel}\label{appB}

A particle residing in cell $j$ experiences a mobile period in the fracture followed by a possible trapping period in the rock matrix. The total residence time in the control volume is
\begin{equation}
\theta_j = \theta_{\mathrm{f},j} + \theta_{\mathrm{m},j} .
\end{equation}
The mobile time is exponential \citep{Russian2016} with mean
\begin{equation}\label{app:eq:tau}
\tau_j = \Big(\sum_{k\in\sigma(j)} b_{kj}\Big)^{-1},
\qquad
\psi_{\mathrm{f},j}(t)=\tau_j^{-1}e^{-t/\tau_j},\quad t>0 .
\end{equation}
Conditioned on a realised mobile duration $r$, the immobile time follows the Lévy--Smirnov first--passage law
\begin{equation}\label{app:eq:psi_m_cond}
\psi_{\mathrm{m},j}(t\mid r)
=
\frac{B_j r}{2\sqrt{\pi}}\, t^{-3/2}
\exp\!\left(-\frac{B_j^2 r^2}{4 t}\right),
\qquad  \text{with~ ~}
B_j = \frac{2\phi_{\mathrm{m}}\sqrt{\alpha_{\mathrm{m}}}}{a_j}.
\end{equation}
Conditional convolution yields the total Laplace transform
\begin{equation}\label{app:eq:psi_total_lap}
\tilde\psi_j(s)
=
\frac{1}{1+\tau_j s + A_j \sqrt{s}},
\qquad \text{with~ ~}
A_j = \tau_j B_j ,
\end{equation}
which reduces to $(1+\tau_j s)^{-1}$ as $A_j\to0$. 

Within the continuous-time random walk (CTRW) formalism, the waiting-time distribution $\psi_j(t)$ uniquely determines the memory kernel $\mathcal K_j(t)$ governing the corresponding generalized master equation (GME). In Laplace space, this relationship reads~\citep[e.g.][]{Dentz2004}
\begin{equation}
\tilde{\mathcal K}_j(s)=\frac{s\tilde\psi_j(s)}{1-\tilde\psi_j(s)}
\end{equation}
The kernel $\mathcal K_j(t)$ represents the non-local-in-time rate at which probability (or energy) leaves cell $j$ due to both advective transport and matrix trapping, and allows the stochastic TDRW formulation to be written in the form of a generalized master equation with memory.

Substituting the Laplace transform of the total residence-time distribution $\tilde\psi_j(s)$ from Eq.~\eqref{app:eq:psi_total_lap} into the CTRW relation above yields the explicit expression
\begin{equation}\label{app:eq:K_total_lap}
\tilde{\mathcal K}_j(s)=\frac{\sqrt{s}}{A_j+\tau_j\sqrt{s}} .
\end{equation}
The inverse Laplace transform of Eq.~\eqref{app:eq:K_total_lap} yields a natural
decomposition of the memory kernel into an instantaneous (Markovian) contribution
and a regular hereditary term,
\begin{equation}\label{app:eq:K_total_split}
\mathcal{K}_j(t)
=
\tau_j^{-1}\,\delta(t)
+
\mathcal{K}_{j,\mathrm{m}}(t).
\end{equation}
The Dirac term $\tau_j^{-1}\delta(t)$ represents the instantaneous mobile
contribution associated with mobile transport within the fracture,
while the second term accounts for delayed release due to matrix trapping.
The latter follows from the regular part of the inverse Laplace transform and
reads
\begin{equation}\label{app:eq:Km}
\mathcal{K}_{j,\mathrm{m}}(t)
=
-\frac{A_j}{\tau_j^{2}\sqrt{\pi t}}
+
\frac{A_j^{2}}{\tau_j^{3}}
\exp\!\left(\frac{A_j^2}{\tau_j^2} t\right)
\operatorname{erfc}\!\left(\frac{A_j}{\tau_j}\sqrt{t}\right).
\end{equation}

The regular part of the kernel, $\mathcal{K}_{j,\mathrm{m}}(t)$, exhibits two distinct asymptotic regimes that reflect the diffusive nature of matrix exchange.

For short times $t\to0^+$, the argument of the complementary error function is small and $\operatorname{erfc}(z)\to1$. The exponential term tends to unity as $t\to0^+$, while the first term in Eq.~\eqref{app:eq:Km} diverges as $t^{-1/2}$. Hence,
\begin{equation}
\mathcal{K}_{j,\mathrm{m}}(t)
\sim
-\frac{A_j}{\tau_j^{2}\sqrt{\pi}}\,t^{-1/2},
\qquad t\to0^+,
\end{equation}
which corresponds to the characteristic $t^{-1/2}$ singularity associated with semi-infinite diffusion.

For long times $t\to\infty$, let's define
\begin{equation}
\mathscr{Z}=\frac{A_j}{\tau_j}\sqrt{t}.
\end{equation}
Using the classical large-argument expansion
\begin{equation}
\operatorname{erfc}(\mathscr{Z})
\sim
\frac{e^{-\mathscr{Z}^2}}{\sqrt{\pi}\mathscr{Z}}
\left(
1-\frac{1}{2\mathscr{Z}^2}+\mathscr{O}(\mathscr{Z}^{-4})
\right),
\qquad \mathscr{Z}\to\infty,
\end{equation}
one obtains
\begin{equation}
\exp\!\left(\frac{A_j^2}{\tau_j^2}t\right)
\operatorname{erfc}\!\left(\frac{A_j}{\tau_j}\sqrt{t}\right)
\sim
\frac{\tau_j}{A_j\sqrt{\pi}}\,t^{-1/2}
-
\frac{\tau_j^3}{2A_j^3\sqrt{\pi}}\,t^{-3/2}
+\mathscr{O}(t^{-5/2}).
\end{equation}

Substituting this expansion into Eq.~\eqref{app:eq:Km} shows that the leading
$t^{-1/2}$ contribution cancels exactly with the explicit $t^{-1/2}$ term already present in the kernel. The first non-vanishing contribution therefore scales as
\begin{equation}
\mathcal{K}_{j,\mathrm{m}}(t)
\sim
-\frac{1}{2A_j\sqrt{\pi}}\,t^{-3/2},
\qquad t\to\infty.
\end{equation}

Thus, while the kernel exhibits a $t^{-1/2}$ singularity at short times,
its long-time tail decays as $t^{-3/2}$, consistently with the first-passage-time scaling of one-dimensional diffusion into a semi-infinite matrix. This cancellation reflects the balance between instantaneous mobile release and delayed matrix return, leaving diffusion-controlled algebraic memory at long times.

Using Equation~\eqref{eq:transit_prob_time} and retaining the Markovian part inside the fracture flux term leads to
\begin{equation}\label{app:eq:GME_separated}
\frac{d g_i(t)}{dt}
=
\sum_{j\in\sigma(i)} \frac{b_{ij}}{V_i} g_j(t)
-
\frac{1}{V_i\tau_i} g_i(t)
-
\int_0^t \mathcal{K}_{i,\mathrm{m}}(t-\tau)\,g_i(\tau)\,d\tau .
\end{equation}
The first two terms constitute the finite--volume (FV) discretisation of the depth--averaged fracture ADE.  
The final term is a causal Volterra sink:
\begin{equation}\label{app:eq:B_def}
\mathcal B_i(t)
=
\int_0^t \mathcal{K}_{i,\mathrm{m}}(t-\tau)\,g_i(\tau)\,d\tau.
\end{equation}
As $A_j\to0$ (no matrix exchange), \eqref{app:eq:GME_separated} reduces exactly to the FV--ADE in the fracture.

Finally, substituting the total kernel into both inflow and outflow operators yields
\begin{equation}\label{app:eq:GME_compact}
\frac{d g_i(t)}{dt}
=
\sum_{j\in\sigma(i)}
\int_0^t\!\mathcal K_j(t-\tau)\,\frac{b_{ij}}{V_i} g_j(\tau)\,d\tau
-
\int_0^t\!\mathcal K_i(t-\tau)\,g_i(\tau)\,d\tau,
\end{equation}
which is algebraically equivalent to \eqref{app:eq:GME_separated}.  
The separated form highlights the FV fracture operator plus a hereditary matrix sink;  
the compact form emphasises the CTRW kernel structure.


\section{Numerical implementation of the TDRW framework}\label{appC}

The numerical implementation follows the workflow illustrated in figure~\ref{fig:fig8}. 
A self–affine aperture field is first generated using the spectral synthesis procedure of \citet{Lenci2022a}, and the corresponding pressure and velocity fields are obtained by solving the Reynolds lubrication equation on the rough fracture geometry. These outputs define the local advective and diffusive fluxes across cell faces, from which jump probabilities and mean advective residence times are assembled.

Thermal particles are then propagated by a Time–Domain Random Walk (TDRW), in which each step comprises an exponentially distributed mobile residence time and a matrix–trapping time drawn from the conditioned L\'evy--Smirnov law derived in Appendix~\ref{appA}. Macroscopic observables, including the first and second longitudinal moments, breakthrough curves, and fracture-matrix heat flux are accumulated on the fly during particle propagation. Matrix heat exchange is evaluated through an event-driven Volterra convolution, requiring only the storage of thermal arrival times and preserving the exact~$t^{-1/2}$ asymptotic behaviour of semi–infinite diffusion.

For clarity and reproducibility, the full computational procedure is reported below in compact pseudocode form. These algorithmic listings complement the description above and provide a direct, actionable summary of the implementation consistent with the modelling framework.

\begin{algorithm}[H]
\caption{\textsc{Main\_Simulation}}
\begin{algorithmic}[1]
  \State $a(\boldsymbol x) \gets$ \textsc{Aperture\_Field\_Module}()
  \State $(p,u_1,u_2) \gets$ \textsc{Reynolds\_Equation\_Module}$(a)$
  \State $(b_{ij},\tau_j,\mathbb P_{j\to i}) \gets$ \textsc{Assemble\_Local\_Transport}$(a,u_1,u_2)$
  \For{each particle}
    \State Initialise at inlet; set $t\gets0$
    \While{particle inside domain}
      \State $(j,t) \gets$ \textsc{TDRW\_Module}$(j,t)$
      \State \textsc{Dynamic\_Transport\_Statistics\_and\_Metrics\_Module}$(j,t)$
    \EndWhile
  \EndFor
\end{algorithmic}
\end{algorithm}

\begin{algorithm}[H]
\begin{algorithmic}[1]
\For{each cell $j$}
  \State Identify neighbours $i\in\sigma(j)$
  \State Compute $b_{ij}$ (upwind advection + in–plane diffusion)
  \State $\tau_j \gets (\sum_{i\in\sigma(j)} b_{ij})^{-1}$
  \State $\mathbb P_{j\to i} \gets b_{ij}\,\tau_j$
\EndFor
\end{algorithmic}
\caption{\textsc{Assemble\_Local\_Transport}}
\end{algorithm}

\begin{algorithm}[H]
\begin{algorithmic}[1]
\State Sample $\theta_{\mathrm{f}}\sim\mathrm{Exp}(\tau_j)$
\State Sample $\theta_{\mathrm{m}}$ from conditioned L\'evy–Smirnov
\State $t\gets t+(\theta_{\mathrm{f}}+\theta_{\mathrm{m}})$
\State Draw next cell $i$ with probability $\mathbb P_{j\to i}$
\If{exit plane reached} record exit time and terminate
\Else $j\gets i$
\EndIf
\end{algorithmic}
\caption{\textsc{TDRW\_Module}}
\end{algorithm}

Macroscopic observables are accumulated on the fly at prescribed observation times $\{t_n\}$,
without storing particle trajectories.
The first and second longitudinal moments are evaluated as
\begin{equation}
\mathcal M(t_n)=\langle x_1\rangle,\qquad 
\mathcal V(t_n)=\langle x_1^2\rangle - \mathcal M^2,
\end{equation}
and exit times increment the complementary $CDF$ to construct $BTC$.

Matrix heat exchange is computed in an event–driven manner.
Each time a particle enters a cell, we log an instantaneous thermal impulse at time $\tau_k$.
At an observation time $t_n$, the local fracture–matrix heat flux is obtained as
\begin{equation}
q_{\mathrm{m}}(\mathbf{x},t_n)=
\frac{2\,\phi_{\mathrm{m}}\,k_{\mathrm{r}}\,(T_1-T_0)}{\sqrt{\pi\,\alpha_{\mathrm{m}}}}
\sum_{\tau_k<t_n}\frac{1}{\sqrt{t_n-\tau_k}},
\end{equation}
which is the discrete form of the Duhamel/Volterra convolution for semi–infinite conduction.
This update reproduces the exact $t^{-1/2}$ memory kernel and requires memory that scales only
with the number of thermal events, not with simulation time or grid size.

Algorithmically, the same physics can be expressed as a continuous–time random walk:  
when a particle experiences an advective residence time $\theta_{\mathrm{f}}$ in a cell, the associated
matrix trapping time is sampled from the L\'evy--Smirnov distribution conditioned on $\theta_{\mathrm{f}}$.
This waiting–time sampling is the stochastic counterpart of the convolution above.
In our implementation we adopt the event–driven impulse summation, while the
waiting–time formulation provides an equivalent CTRW representation; both routes yield identical
breakthrough curves and reproduce the characteristic $t^{-1/2}$ long–time behaviour.

\subsection{Computational cost and scaling}
The computational cost scales linearly with the number of jump events 
$N_{\mathrm{ev}}\sim N_{\mathrm{p}}\,\chi (L/\Delta x)$, where the Eulerian tortuosity 
$\chi = \langle u_1\rangle/\langle\|\boldsymbol{u}\|\rangle$ (computed from the depth–averaged velocity field excluding near–contact cells) quantifies streamline elongation. 
For the strongly heterogeneous fracture 
($L/L_\mathrm{c} = 2^5$, closure = 1), we obtain 
$\chi\simeq 1.35$, so each particle performs on average 
$\chi (L/\Delta x)\approx 1.4\times 10^3$ advective jumps on a $2^{10}$–cell mesh. 
On a workstation equipped with an AMD Ryzen 9 7940HX (16 cores, 32~GB RAM), a simulation with 
$10^6$ particles ($\approx 1.4\times 10^9$ events) required 
$3.2\times 10^3$~s ($\approx 4.3\times 10^5$ events~s$^{-1}$, 
$\approx 2.7\times 10^4$ per core), while $10^5$ particles completed in $\approx 3\times 10^2$~s, confirming near–linear scaling ($\approx 6$–7\% overhead when increasing problem size by a factor~10). 
The event–driven Volterra update stores only arrival times, so memory and CPU grow with thermal events rather than wall time or grid size, keeping the method tractable for large meshes and ensembles.

\bibliographystyle{jfm}
\bibliography{references}

\end{document}